\def\spose#1{\hbox to 0pt{#1\hss}}

\def\multleft#1{\hbox to size{\vbox {\halign {\lft{##}\cr #1}}\hfill}\par}
\def\multright#1{\hbox to size{\vbox {\halign {\rt{##}\cr #1}}\hfill}\par}

\def\today{\ifcase\month\or January\or February\or March\or April\or May\or
      June\or July\or August\or September\or October\or November\or December\fi
      \space\number\day, \number\year}
\def\s{\hbox{\phantom{5}}}      

\def\erg{{\rm\thinspace erg}}

\def\mJy{{\rm\thinspace mJy}}
\def\microJy{{\rm\thinspace $\mu$Jy}}
\def\micron{{\rm\thinspace $\mu$m}}

\def\K{{\rm\thinspace K}}
\def\keV{{\rm\thinspace keV}}

\def\Lsun{\hbox{$\rm\thinspace L_{\odot}$}}

\def\Msun{\hbox{$\rm\thinspace M_{\odot}$}}

\def\s{{\rm\thinspace s}}

\def\yr{{\rm\thinspace yr}}

\def\ergps{\hbox{$\erg\s^{-1}\,$}}

\def\Msunpyr{\hbox{$\Msun\yr^{-1}\,$}}

\def\H2{\hbox{H$_{2}$}}

\documentclass[usegraphicx]{mn2e}

\usepackage{times}
\usepackage{amssymb}
\usepackage{eso-pic}
\include{defn}

\begin{document}
\hsize=6truein
\title[An infrared--radio simulation of the extragalactic sky]
{An infrared--radio simulation of the extragalactic sky: from the Square Kilometer Array to Herschel}
\author[R.J.~Wilman et al.]
{\parbox[]{6.in} {R.J.~Wilman$^{1}$, M.J.~Jarvis$^{2}$, T.~Mauch$^{3}$, S.~Rawlings$^{3}$ and S.Hickey$^{2}$ \\ \\
\footnotesize
1. Centre for Astrophysics \& Supercomputing, Swinburne University of Technology, Hawthorn, Victoria 3122, Australia \\
2. Centre for Astrophysics, Science \& Technology Research Institute, University of Hertfordshire, Hatfield, AL10 9AB\\ 
3. Oxford Astrophysics, Denys Wilkinson Building, Keble Rd, Oxford, OX1 3RH\\ }}
\maketitle

\begin{abstract}
To exploit synergies between the {\em Herschel} Space Observatory and next generation radio facilities, we have extended 
the semi-empirical extragalactic radio continuum simulation of Wilman et al. to the mid- and far-infrared. 
Here we describe the assignment of infrared spectral energy distributions (SEDs) to the star-forming galaxies and active galactic nuclei,
using {\em Spitzer} 24, 70 and 160\micron~and {\em SCUBA} 850\micron~survey results as the main constraints.

Star-forming galaxies dominate the source counts, and a model in which their far-infrared--radio correlation and infrared SED assignment procedure are invariant with redshift 
underpredicts the observed 24 and 70\micron~source counts. The 70\micron~deficit can be eliminated if the star-forming galaxies undergo stronger luminosity evolution than 
originally assumed for the radio simulation, a requirement which may be partially ascribed to known non-linearity in the far-infrared--radio correlation at low luminosity if it
evolves with redshift. At 24\micron, the shortfall is reduced if the star-forming galaxies develop SEDs with cooler dust and correspondingly stronger Polycyclic Aromatic Hydrocarbon (PAH) emission features with increasing redshift at a 
given far-infrared luminosity, but this trend may reverse at $z>1$ in order not to overproduce the sub-mm source counts. The resulting model compares favourably with recent {\em BLAST} results and we have extended the 
simulation database to aid the interpretation of {\em Herschel} surveys. Such comparisons may also facilitate further model refinement and revised predictions for the {\em Square Kilometer Array} and its precursors.
\end{abstract}


\section{INTRODUCTION}
In Wilman et al.~(2008) (hereafter W08) we presented a semi-empirical simulation of the extragalactic radio continuum sky primarily intended to aid the design of 
scientific programmes for the next generation of radio telescope facilities, culminating in the {\em Square Kilometer Array} (SKA). 
The simulation covers a sky area of $20 \times 20$~deg$^{2}$ and contains radio-loud and radio-quiet active galactic nuclei (AGN) and star-forming galaxies out 
to redshift $z=20$ within a framework for their large-scale clustering. The full source catalogue -- containing 320 million sources above 10~nJy at five frequencies ranging from 151~MHz to 18~GHz -- 
can be accessed at the SKADS Simulated Skies ($S^{3}$) database\footnote{http://s-cubed.physics.ox.ac.uk} under $S^{3}$--SEX (semi-empirical extragalactic simulation).

There are by necessity numerous uncertainties and limitations in the $S^{3}$--SEX simulation, including but not limited to issues such as the form of the high-redshift evolution of the AGN and galaxies, 
the lack of star-forming/AGN hybrid galaxies, and the abundance of highly-obscured Compton-thick AGN. In so far as possible, flexibility was built into the simulations to allow post-processing to improve their accuracy as observations in the years ahead lead to improved constraints. Major advances in these areas are expected from the far-infrared surveys to be performed by the {\em Herschel} Space Observatory (Pilbratt 2008). To facilitate such comparisons we have post-processed the $S^{3}$--SEX simulation to cover these wavelengths. In this paper, we describe the recipes for assigning infrared spectral energy distributions (SEDs) to the radio sources, using existing mid- and far-infrared results from {\em Spitzer} and sub-mm survey data from {\em SCUBA} as the primary constraints. We then present our predictions for {\em Herschel} surveys, bolstered by a comparison with results from the 
Balloon-borne Large Aperture Submillimetre Telescope ({\em BLAST}) which offer a foretaste of {\em Herschel}'s capabilities. In keeping with the philosophy of the S$^{3}$ project to maximise the degree of interaction between the user and the database, the infrared fluxes are also provided on the $S^{3}$ webpage from which users can generate simulation products for comparison with observations. The radio--infrared connection is of immense empirical 
and theoretical interest for extragalactic surveys, for the identification and follow-up of {\em Herschel} sources, and for probing the physics of the far-infrared--radio correlation and its possible evolution. The simulation
can nevertheless also serve as a standalone {\em Herschel} simulation, for comparison with others such as the phenomenologically-inspired backward evolution models of Valiante et al.~(2009) and Pearson \& Khan~(2009), and the model of Lacey et al.~(2009) which is based on a semi-analytical galaxy formation model.

\section{THE ASSIGNMENT OF INFRARED SEDs TO THE RADIO SOURCE POPULATIONS}
The input radio source catalogue was obtained from the $S^{3}$-SEX online database and negative evolution at high 
redshift was imposed using the `default post-processing options' described in W08. Each radio source was assigned an 
infrared SED from various template libraries appropriate for star-forming galaxies and AGN, and the output flux density of
each infrared model component for each galaxy [in log(F$_{\rm{\nu}}$(Jy)] was computed at observed wavelengths of 24, 70, 100, 160, 250, 350, 450, 500, 850, 
and 1200 \micron. The wavelengths of 24, 70 and 160 \micron~are those of the {\em Spitzer} MIPS instrument, for which an abundance of extragalactic survey 
data is available to guide the construction of our model and to compare against its output (see the review of Soifer et al.~2008). {\em Herschel} will conduct surveys at 250, 350 and 500 \micron~with the SPIRE instrument and at 70, 100 and 160\micron~with PACS. Longer-wavelength constraints are available from submillimetre surveys by SCUBA at 450 and 850\micron, and from the MAMBO instrument on the 30-metre IRAM telescope at 1200\micron. The catalogued output flux densities are monochromatic values, with the exception of the {\em Spitzer} 24 and 70\micron~MIPS bands for which the modelled spectra were convolved with the MIPS bandpass transmission curves, following the {\em Spitzer} Synthetic Photometry Recipe\footnote{http://ssc.spitzer.caltech.edu/postbcd/cookbooks/synthetic\underline{ }photometry.html}.  This is due to the spectral complexity in the rest-frame 10\micron~region, particularly for the star 
formation component due to the presence of Polycyclic Aromatic Hydrocarbon (PAH) features. 

The format of the output from the infrared extension, as it appears in the $S^{3}$-SEX online database, is described in Appendix A.

\subsection{Star-forming galaxies}
The population of star-forming galaxies in W08 comprises two populations, normal (or quiescent) galaxies and starbursts, identified respectively with the low and high-luminosity Schechter function components of the local 1.4~GHz luminosity function of Yun, Reddy \& Condon (2001). The entire population was evolved with pure luminosity evolution (1+z)$^{3.1}$ out to z=1.5 (defined in an Einstein-de Sitter cosmology but adapted to the flat-$\Lambda$ cosmology used for the simulation). The default post-processing option consists of negative space density evolution of the form $(1+z)^{-7.9}$ above $z=4.8$. We stress that the terms `normal galaxy' and `starburst' are merely convenient labels for these two components of the local luminosity function in our phenomenological model. Physically speaking, the terms may not necessarily offer an accurate description of the nature of the star formation in these populations, especially beyond the local Universe.

The first step in assigning an infrared SED is to use the far-infrared--radio correlation for
star-forming galaxies (eqn.~5 of Yun, Reddy \& Condon~2001) in order to derive the rest-frame far-infrared luminosity, L(FIR), given the rest-frame 1.4~GHz luminosity, L$_{\rm{1.4 GHz}}$. The relation is characterised by the ``q'' parameter:

\begin{equation}
q = \rm{ log [L(FIR,W)}/3.75 \times 10^{12} Hz] - log [\rm{L_{\rm{1.4 GHz}} (W Hz^{-1})}],
\end{equation}

for which we assume a gaussian distribution with $\mu = 2.34$ and $\sigma = 0.26$ (Yun, Reddy \& Condon). There is evidence that this relation holds out to redshift $z > 1$ (e.g. Appleton et al.~2004, Garn et al.~2009b), 
although such conclusions are sensitive to the assumed infrared SEDs and associated ``k-corrections''. Indeed, using recent {\em BLAST} stacking measurements, Ivison et al.~(2009) reported evidence for a tentative decline in $q_{\rm{IR}}$ (defined as the logarithmic ratio of the the bolometric 8--1000\micron~infrared to monochromatic 1.4~GHz radio fluxes), of the form $q_{\rm{IR}} \propto (1+z)^{-0.15 \pm 0.03}$ (see also Kovacs et al.~2006).

The far-infrared luminosity, L(FIR), is traditionally defined through a linear combination of IRAS 60 and 100\micron~flux densities~(see Sanders \& Mirabel~1996), which for an SED
consisting of a superposition of modified blackbodies yields the 42.5--122.5\micron~luminosity to sub-percent accuracy (Helou et al.~1988). Given L(FIR), 
galaxies were assigned a model SED from the library of star-forming galaxy templates assembled by Rieke et al.~(2009). The latter consists of 14 SEDs uniformly spaced in total 
infrared luminosity (8--1000\micron), L(TIR), from log L(TIR)(\Lsun) = 9.75 to 13.00. The templates were taken from the online version of Table 4 in Rieke et al.~(2009) but they 
terminate on the short wavelength side at 5.26\micron~for those with log L(TIR) (\Lsun) $\leq 11$ and at 4.02\micron~at higher luminosities. In order to predict
24\micron~fluxes for galaxies at redshift $z \geq 3$, we need to extend these templates to shorter wavelengths. For this purpose we employed the spectra of the individual local luminous
(LIRGs) and ultraluminous infrared galaxies (ULIRGs) in Table 3 of Rieke et al.~(2009), which extend down to 0.4\micron. From these, we computed mean LIRG and ULIRG spectra and matched
them onto the original Rieke templates below the cut-off wavelengths of 4.02 and 5.26\micron. The ULIRG spectrum was used for the templates with log L(TIR)(\Lsun) $>$ 12, and the 
LIRG spectrum for the remaining templates. The resulting templates should be considered as schematic below $\sim 3$\micron, as we did not attempt to model the stellar population and rest-frame optical 
extinction in a self-consistent fashion.

\begin{figure}
\includegraphics[width=0.47\textwidth,angle=0]{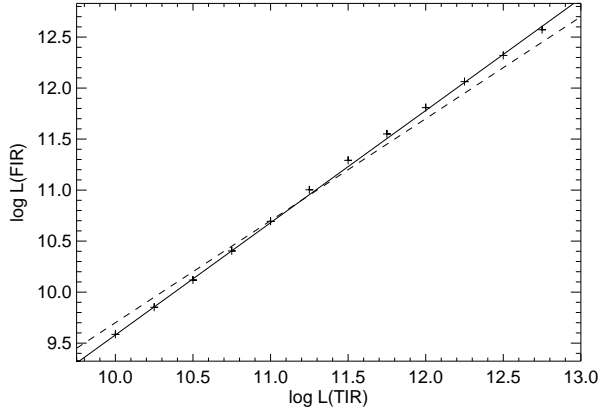}
\caption{\normalsize  The symbols show the far-infrared luminosity (42.5--122.5~\micron), L(FIR)(\Lsun), of the star-forming galaxy templates of Rieke et al.~(2009) as a function of 
total infrared luminosity, L(TIR). To assign SED templates to the simulated galaxies, we assume the non-linear relation given by the solid line: log L(FIR) = 1.1log L(TIR) -- 1.42. The 
dashed line shows the linear relation L(FIR) = 1.91 L(TIR) assumed by Marcillac et al.~(2006).}
\label{fig:LTIRtoFIR_Rieke}
\end{figure}

\begin{figure}
\includegraphics[width=0.47\textwidth,angle=0]{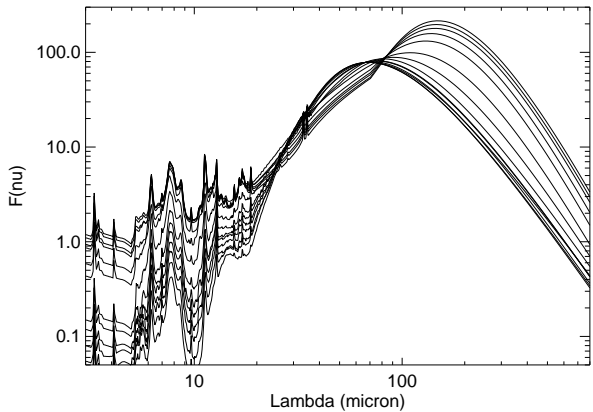}
\caption{\normalsize  The star-forming galaxy template SEDs scaled to a common far-infrared luminosity using the L(TIR)--L(FIR) scaling in Fig.~\ref{fig:LTIRtoFIR_Rieke}. With increasing
L(TIR), the SEDs peak at shorter wavelengths and exhibit weaker spectral features in the $\lambda <20$\micron~region.}
\label{fig:RiekesameFIR}
\end{figure}

The template SEDs were integrated over the appropriate wavelengths ranges to yield the mapping between
L(TIR) to L(FIR) shown in Fig.~\ref{fig:LTIRtoFIR_Rieke}; the adopted relation between the two quantities is mildly non-linear: log L(FIR) = 1.1log L(TIR) -- 1.42. 
Based on the IRAS Bright Galaxy Sample (Soifer et al.~2007), Marcillac et al.~(2006) assumed a linear relation: L(FIR) = 1.91 L(TIR). In Fig.~\ref{fig:RiekesameFIR} we show the Rieke 
templates scaled to a common L(FIR).

\begin{figure}
\includegraphics[width=0.47\textwidth,angle=0]{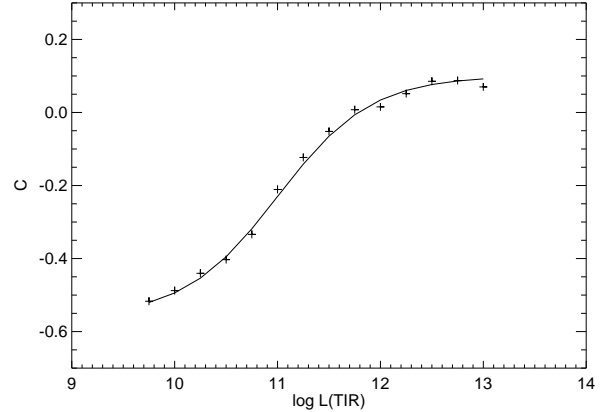}
\caption{\normalsize Colour of the Rieke templates [$C = \rm{log} L_{\rm{60}}/L_{\rm{100}}$] as a function of log L(TIR)(\Lsun), with the fit given by equation (2) of section 2.1.}
\label{fig:RiekeCOLOR}
\end{figure}

At a given L(FIR), the local galaxy population exhibits a distribution in 60--100\micron~colour or, equivalently, dust temperature, as characterised by Chapin et al.~(2009). We use this 
distribution (Chapin et al.~2009, equations 8 and 9) to assign each star-forming galaxy a 60--100\micron~colour, $C$, defined as the ratio of the rest-frame monochromatic 60 and 100\micron~flux densities: 
$C = \rm{log} L_{\rm{60}}/L_{\rm{100}}$. The template nearest in colour is then selected from the Rieke library using the empirical $C$--L(TIR) conversion for these templates shown in Fig.~\ref{fig:RiekeCOLOR}, for
which a functional fit is:

\begin{equation}
C = 0.33 \rm{tanh}[1.1(\rm{log L(TIR)} - 11.0)] - 0.23.
\end{equation} 

Having selected the template shape from the Rieke library by this process, the final step is to normalise it to actual value of L(FIR) originally specified by the far-infrared--radio relation. This is done using the non-linear L(FIR)--L(TIR) relation given in Fig.~\ref{fig:LTIRtoFIR_Rieke}. Chaplin et al.~(2009) presented evidence for evolution 
in the colour-luminosity relation, but we initially assume that the local relation applies at all redshifts.

\subsection{Radio-quiet AGN}
\begin{figure*}
\includegraphics[width=0.35\textwidth,angle=0]{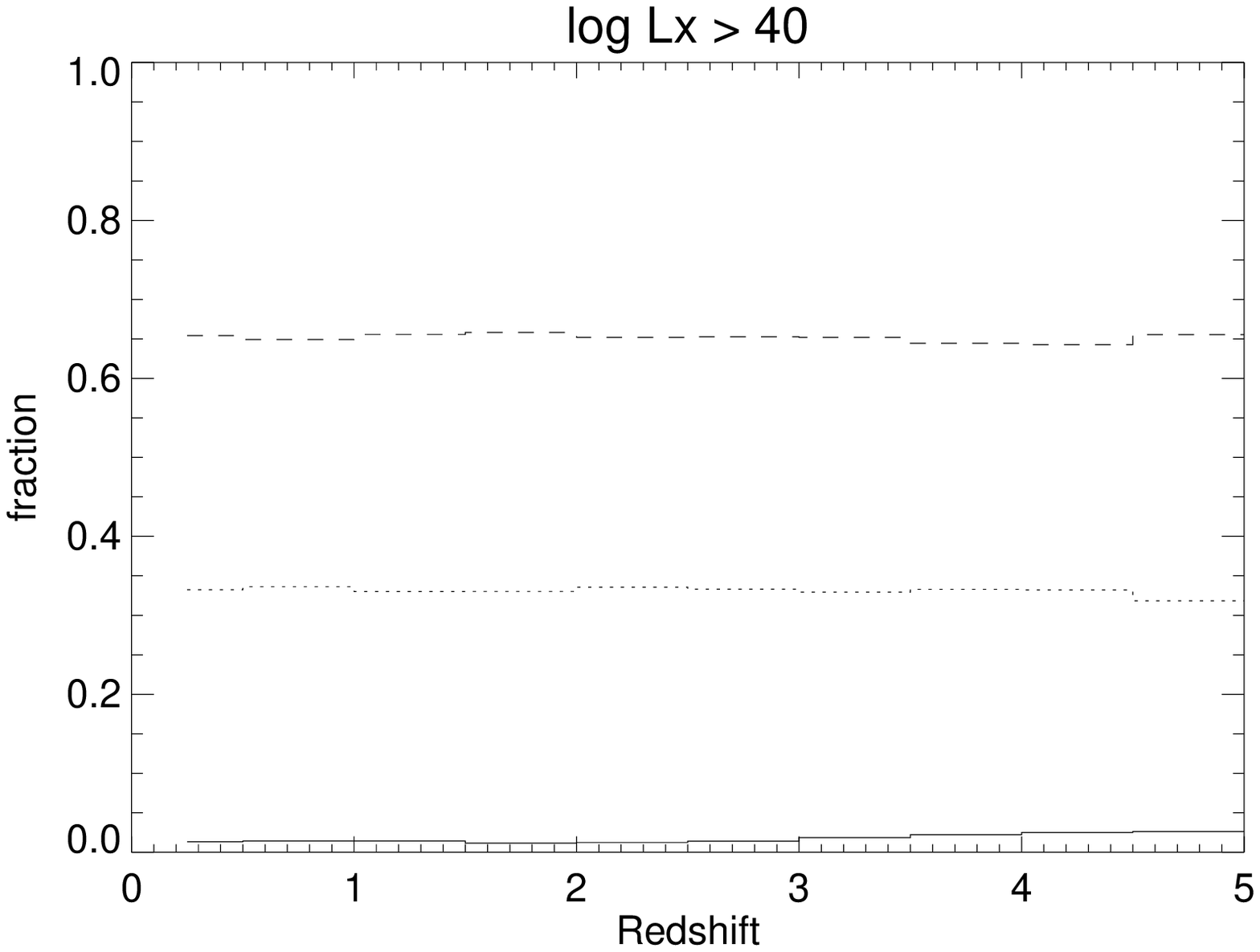}
\includegraphics[width=0.35\textwidth,angle=0]{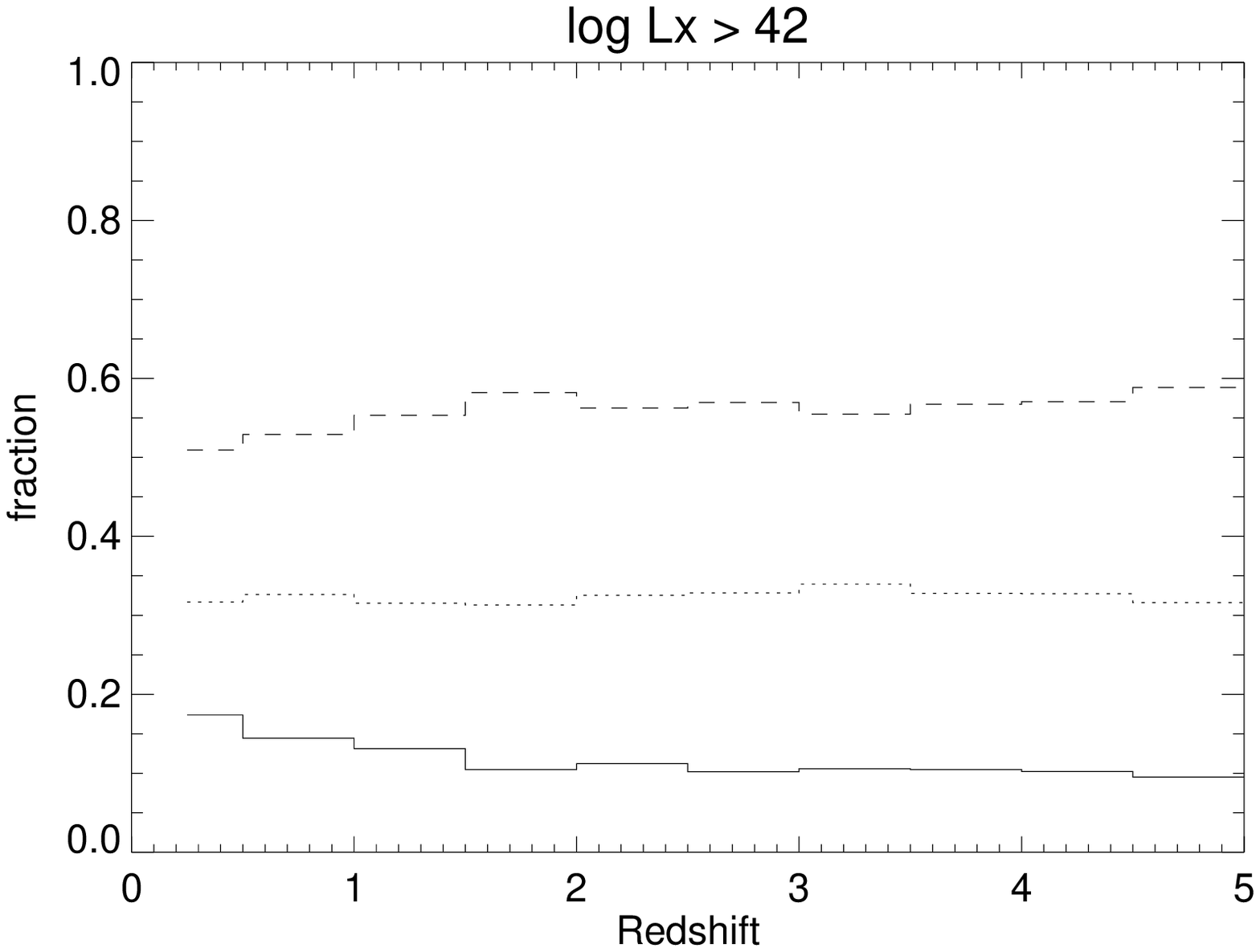}
\includegraphics[width=0.35\textwidth,angle=0]{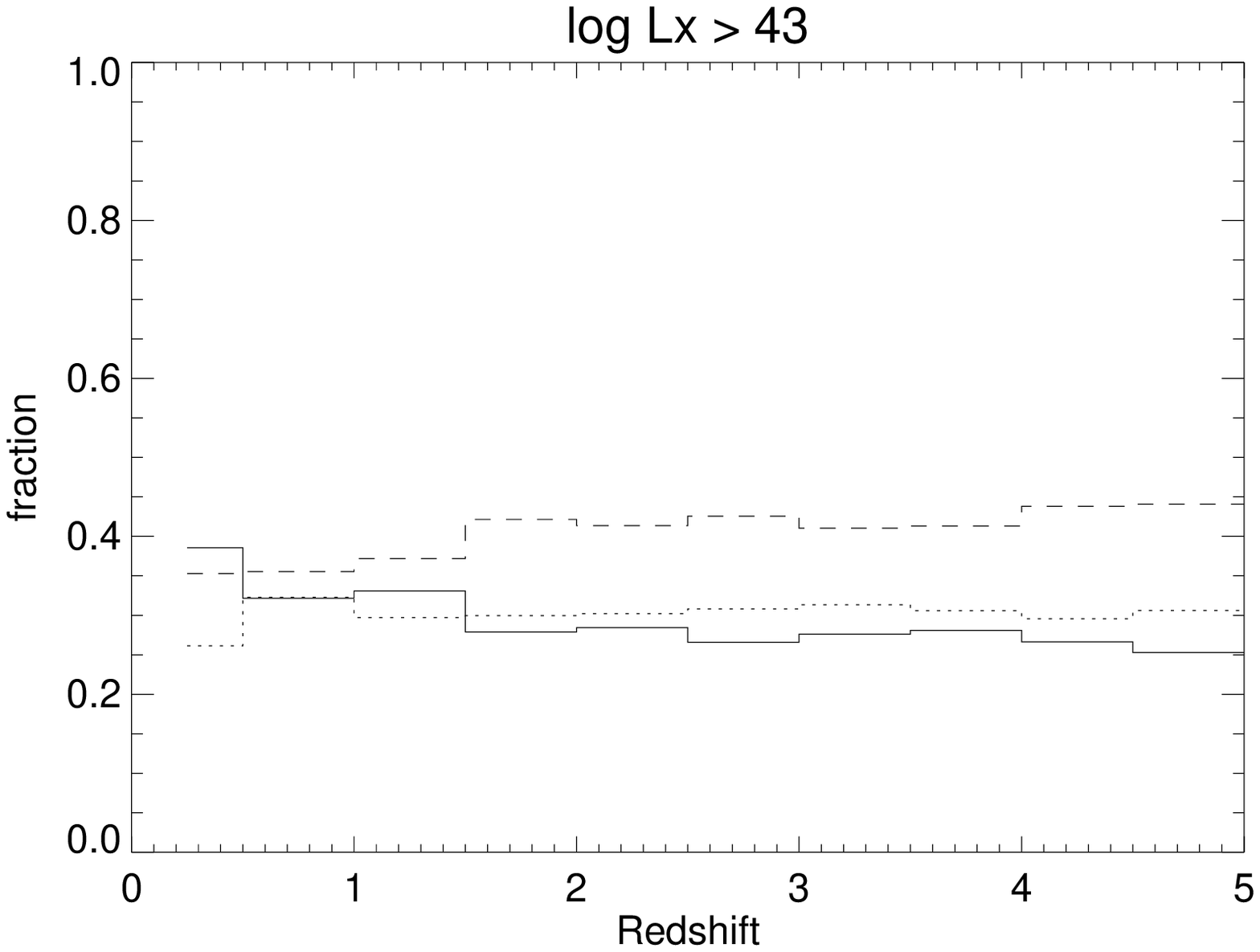}
\includegraphics[width=0.35\textwidth,angle=0]{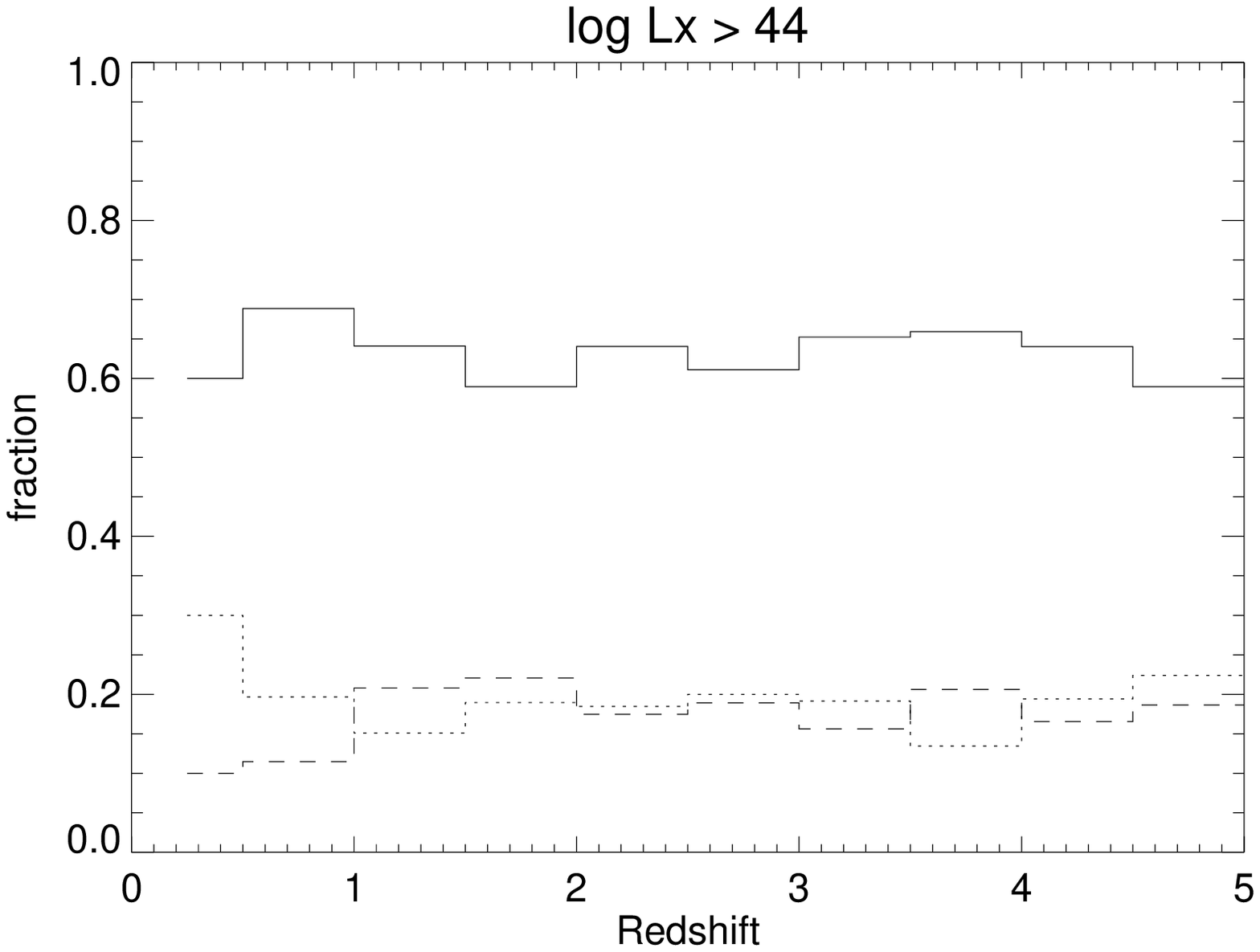}
\caption{\normalsize The fractions of retained radio-quiet AGN identified as unobscured 
AGN (solid lines), Compton-thin obscured (dashed lines) and Compton-thick obscured AGN (dotted lines), for intrinsic luminosity thresholds of $L_{\rm{X}}> 10^{40}$, $10^{42}$, $10^{43}$ and $10^{44}$\ergps.
Evaluated from a 4~deg$^{2}$ simulation area.}
\label{fig:RQAGNfrac}
\end{figure*}

In W08 a population of radio-quiet AGN was incorporated using the Ueda et al.~(2003) hard X-ray luminosity function and a relation between 2--10\keV~and 1.4~GHz 
luminosities. To assign infrared SEDs the first task is to split the population into subgroups of unobscured AGN, and Compton-thin and Compton-thick obscured AGN. This was not done in 
W08 but is now necessary because the infrared SEDs are sensitive to this classification. We start from the findings of Hasinger~(2008), who used a compilation of hard X-ray surveys to parameterise 
the fraction of obscured Compton-thin AGN (as a proportion of a total population comprising unobscured AGN and obscured Compton-thin AGN, but excluding Compton-thick obscured AGN) 
as a function of $L_{\rm{X}}$ (2--10\keV~luminosity; \ergps):

\begin{equation}
f_{\rm{2}} = [0.27 + \beta(log L_{\rm{X}} - 43.75)]g(z),
\end{equation}

where $\beta = -0.281$ and $g(z)$ is evolution of the form (1+z)$^{0.62}$ out to $z=2$ and flat thereafter. The number of Compton-thick obscured AGN is not as well 
constrained, and in W08 we simply boosted the space density of the Ueda et al.~(2003) luminosity function by 50~per cent in a notional allowance for them, but left the 
issue open for subsequent refinement in post-processing. Ueda et al. showed that a reasonable match to the X-ray background results if the number of Compton-thick AGN simply equals the
number of Compton-thin obscured AGN. We thus assume that the abundance of Compton-thick AGN population is a factor $f_{\rm{Cthick}}$ times that of the combined population of unobscured and 
Compton-thin obscured AGN, where $f_{\rm{Cthick}}$ = min[0.5,$f_{\rm{2}}$], with the upper bound of 0.50 hard-wired into the W08 simulation. After the removal of any excess Compton-thick AGN, 
the remaining sources in the W08 catalogue are probabilistically identified as unobscured (type I), Compton-thin or Compton-thick obscured (type II) AGN. Finally, 10~per cent of all 
sources are flagged for removal because W08 included radio-loud AGN with a separate radio luminosity function but this population is also implicitly present in the Ueda et al. luminosity function. 
The fractions of retained radio-quiet AGN which are unobscured, Compton-thin and Compton-thick obscured are shown as functions of $L_{\rm{X}}$ and redshift in Fig.~\ref{fig:RQAGNfrac}.

\subsubsection{AGN emission}
For the retained sources, $L_{\rm{X}}$ is inferred from the radio luminosity using equation (2) of W08. Sources with  $L_{\rm{X}}$ above and below a threshold 
of $3 \times 10^{44}$\ergps~are classified as quasars and Seyferts, respectively. Unobscured (type I) quasars are assigned at random one of the three templates for optically-selected
QSOs compiled by Polletta et al.~(2007) (termed QSO1, BQSO1 and TQSO1, which differ in their relative optical/infrared flux ratios)\footnote{These SEDs were obtained from http://www.iasf-milano.inaf.it/$\sim$polletta/templates/swire\underline{ }templates.html}. The unobscured Seyferts are assigned SEDs from the online library of models computed by Nenkova et al.~(2008) using the {\it CLUMPY} code for dust radiative transfer in a clumpy AGN torus. To narrow down the parameter space of 
available models, we use the results of Thompson et al.~(2009) which are based on analysis of Seyfert 1 AGN mid-infrared spectra. They prefer a range of {\it CLUMPY} parameter space defined
by the parameters $Y=30$, $q=1$, $\sigma = 45$~deg, $m=2$, $i=30$~deg, $\tau_{\rm{V}}=30-60$ and $N_{\rm{0}} \leq 6$ (see Nenkova et al.~2008 for the meaning of these parameters). 18 CLUMPY
models satisfy these constraints and were assigned at random to our type I Seyferts, with the SED consisting of the reprocessed torus emission and direct AGN broken power-law emission
(defined as a power-law $F_{\rm{\lambda}} \propto \lambda^{-1.5}$ over 0.1--1\micron~with a Rayleigh-Jeans slope at longer wavelengths). 

For the type II Compton-thin AGN, we assigned one of the type 2 QSO templates given by Polletta et al.~(2006) and Polletta et al.~(2007) (termed QSO2 and Torus, respectively) for the quasars, 
and one of the aforementioned CLUMPY models (but without the direct AGN emission component) to the Seyferts. 

For the Compton-thick AGN, we assigned half the population SEDs using the same prescription as for the type II Compton-thin sources; for the other half of the population, we reflect the growing 
body of evidence which suggests that a large proportion of Compton-thick AGN are obscured by an extended distribution of cold dust in the host galaxy and not by a nuclear torus (see e.g. 
Polletta et al., 2006, 2008; Martinez-Sansigre et al. 2006, Lacy et al. 2007). Accordingly we assign these sources an intrinsic type I or Compton-thin type II SED, extinguished by a 
dust screen with $A_{\rm{V}} = 4-  27$~mag~(as found by Polletta et al.~2008 for obscured quasars). An extinction curve extending into the infrared is taken from McClure~(2009).
These AGN infrared SED components are normalised using the correlation between 12 micron and hard X-ray emission found by Gandhi et al.~(2009), which appears to hold for obscured and unobscured
AGN (including Compton-thick sources) and to extend into the quasar regime:

\begin{equation}
\rm{ log L_{\rm{MIR}}} = (-4.37 \pm 3.08) + (1.106 \pm 0.071)\rm{log} L_{\rm{X}},  
\end{equation}

where L$_{\rm{MIR}} = \lambda L_{\rm{\lambda}}$ at 12.3\micron~in \ergps. Gaussian scatter of $\sigma = 0.36$ is assumed on the relation.

\subsubsection{Star-formation}
There is abundant evidence for ongoing star formation in the host galaxies of radio-quiet AGN and this is likely to dominate the far-infrared SED. The consensus 
emerging from targeted observations of local Seyferts (Buchanan et al.~2006; Thompson et al.~2009) and studies of more distant samples and cosmological surveys (e.g. Silverman et al.~2009; Hern\'{a}n-Caballero et al.~2009; Serjeant \& Hatziminaoglou~2009) is that the star formation is generally higher in obscured AGN compared with their unobscured counterparts, and increases sub-linearly with AGN luminosity. We embody these findings with a star formation rate (SFR) $\propto \sqrt{L_{\rm{X}}} (1+z)^{1.6}$, but assume different 
normalizations for the type I and II AGN. This luminosity and redshift dependence is similar to that measured for quasars by Serjeant \& Hatziminaoglou~(2009) and the luminosity 
dependence matches that of local Seyfert 1s (Thompson et al.~2009). 

For type I AGN, we normalise the relation using the measurements of type I SDSS AGN at z=0.15 by Kim et al.~(2006), which have a median SFR of 0.5\Msunpyr. The median log L([OIII]$\lambda5007$) of this sample, as inferred from Fig.~4(b) of Kim et al., is 41.4 (\ergps) which equates to log L$_{\rm{X}}=42.6$ using the  log L$_{\rm{X}}$--log L([OIII]$\lambda5007$) relation of eqn.~(1) of Silverman et al.~(2009). Thus for type I AGN:

\begin{equation}
{\rm SFR} = 0.63 \sqrt{L_{\rm{X}} /10^{43}} (1+z)^{1.6} \Msunpyr.
\end{equation}

For the type II AGN (Compton-thin and Compton-thick), we base our results on the $z=0.1$ SDSS type II AGN sample of Kauffmann et al.~(2003), focussing on the sub-sample of strong-lined sources with log L([OIII])$>$ 40.5 indicated by the red dots in Fig.~10 of Silverman et al.~(2009). Referring once more to the log L([OIII]) distributions in Fig.~4(b) of Kim et al.~(2006), we infer for this particular luminosity-limited sample a median log L([OIII])=41.1, which translates into log L$_{\rm{X}}=42.25$ and hence boosts the pre-factor of our assumed SFR(L$_{\rm{X}}$,z) relation from 0.63 to 2\Msunpyr. A scatter of 1 dex is assumed on the SFR(L$_{\rm{X}}$,z) relation and the star-formation rates are assumed to decline as $(1+z)^{-7.9}$ at $z > 4.8$ following the assumed evolution of the global star formation 
rate density. SEDs for the star formation component are assigned from the Rieke et al.~(2009) library, using the relation given by Yun, Reddy \& Condon~(2001) (their eqn.~13) to relate the star formation rate and 1.4~GHz radio luminosity; the far-infrared--radio relation (eqn.~1) was then used to assign L(FIR) and hence an appropriate template from the Rieke library using Fig.~\ref{fig:LTIRtoFIR_Rieke}.

\subsection{Radio-loud AGN}
Radio-loud AGN were modelled by W08 using the Willott et al.~(2001) 151~MHz luminosity function comprising a weakly-evolving low luminosity component and a rapidly-evolving high luminosity component, identified with FRI and FRII radio sources, respectively. We consider them separately for the purposes of infrared SED assignment.

\subsubsection{FRIs}
There is scant evidence for the presence of intrinsically luminous optical-UV AGN in the radio population we associate with FRI sources (e.g. Vardoulaki et al.~2008). We therefore model their infrared SEDs with a template representing host galaxy starlight only, which we produced using the GALSYNTH online interface to the GRASIL code developed by Silva et al.~(1998). The models are based on their `Elliptical model', which gives the evolving SED (taking account of dust reprocessing) of a stellar population formed in a burst of 1~Gyr duration (see description of this particular model at http://adlibitum.oat.ts.astro.it/silva/grasil/modlib/modlib.html). We generated models with ages separated by 0.1~Gyr for 0.1--1.5~Gyr, and by 1~Gyr for ages 2--13~Gyr. We assume formation redshifts distributed uniformly in the range z=10--20, and select the template nearest in age at the redshift of the galaxy under consideration. To normalise the SEDs, we draw upon the bivariate 1400~MHz--K-band luminosity 
function derived by Mauch \& Sadler~(2007) for a local sample: given a radio power, the absolute K-band magnitude is drawn from a Gaussian distribution of specified mean and variance and this is used
to normalise the SED. For simplicity, we assume the same (local) relation between K-band luminosity and radio power at all redshifts. An artifact of this choice of model is that the mid-far infrared regions
of the SED template increase sharply for ages below the assumed duration of the starburst (1~Gyr). This causes discontinuities in the mid- and far-infrared fluxes at $z \simeq 4$ where the galaxy age first drops below this figure.

\subsubsection{FRIIs}
The FRIIs are first split into populations of quasars and radio galaxies for which the nuclear emission is seen directly and through obscuration, respectively. Willott et al.~(2000) found that for log L(151~MHz)[W/Hz/sr]$>26.5$ the quasar fraction is 0.4, independent of redshift and luminosity, and around 0.1 at lower radio luminosity. Taking into account the contamination by FRIs at lower luminosity, we simply assume that 
40 per cent of our FRIIs are seen as quasars, i.e. those seen at viewing angle of $\leq 53$~degrees to the jet axis. To assign AGN and starburst infrared SEDs, we recall that quasars are, on average, more intrinsically luminous AGN than the radio galaxies (see e.g. Simpson 2003). To work with a parameter that is closely tied to the intrinsic strength of the nuclear emission, we convert from L(151~MHz) [W/Hz/sr] to L([OIII]$\lambda5007$) [W] using a relation
from Grimes, Rawlings \& Willott~(2004) (GRW04):

\begin{equation}
\rm{log L([OIII]\lambda5007}) = 7.53 + 1.045\rm{log L(151~MHz).}
\end{equation}

Eqn.~(6) applies to the population as a whole, but following GRW04 we assume that the quasars and radio galaxies lie 0.3 dex systemtically above and below it, respectively, with an additional 0.5 dex of scatter on each distribution. To assign an AGN component to the infrared SED, we draw upon the findings of Haas et al.~(2008) who presented {\em Spitzer} 3--24\micron~photometry of 3CR steep-spectrum quasars and radio galaxies at redshift $1<z<2.5$. The quasar mid-infrared SEDs are very uniform, roughly constant in $\nu F_{\rm{\nu}}$. There is a near proportionality between $\nu F_{\rm{\nu}}$(8~\micron, rest) and $\nu F_{\rm{\nu}}$(178~MHz, rest) with 0.5 dex scatter. The radio galaxies are
dominated by host galaxy starlight below 3\micron~(rest) and at longer wavelengths by a quasar nucleus reddened by up to A$_{\rm{V}}=50$~mag. The radio galaxies are a factor 3--10 less luminous in this spectral range than the 
quasars, which they attribute entirely to extinction. For a sample of 3CR sources at $0.4 \leq z <1.2$, Cleary et al.~(2007) find that the quasars are 4 times as luminous as the radio galaxies at 15\micron, half of which they 
attribute to extinction, and half to synchrotron contamination in the quasars. However, their sample includes a significant number of steep-spectrum compact and super-luminal quasars. They also assumed a spherically symmetric 
dust model, which is likely to be inadequate.

We assume that the quasars and radio galaxies follow the same intrinsic relation between $\nu L_{\rm{\nu}}$(8~\micron~rest) [W] and L([OIII]$\lambda5007$) [W], derived from the quasar points in Fig.~1 of Haas et al.~(2008) using eqn.~(6) to convert from L(151~MHz) to L([OIII]$\lambda5007$). We assume:

\begin{equation}
\rm{log \nu L_{\rm{\nu}} (8~\micron~rest)} = 3.3 + 
0.9569\rm{log L([OIII]\lambda5007}). 
\end{equation}

No intrinsic scatter is assumed on this, as the scatter in eqn.~(6) can reproduce the scatter in Fig.~1 of Haas et al.~(2008). Using this normalisation, we assign the quasars randomly to one of the three type I quasar templates from
Polletta et al.~(2007). The radio galaxies are also assigned one of these template SEDs in an analogous manner, and then extinction of $A_{\rm{V}}$=0--40~mag is applied. An elliptical host galaxy model from the GRASIL library is assigned as for the FRIs but in this case normalised using the K--z relation of Willott et al.~(2003) with an additional offset of $\Delta K = -0.36(\rm{log L(151 MHz)[W/Hz/sr]} - 26.55)$~mag (following Willott et al.~2003, Fig.~3).

In principle, the evolving elliptical template from the GRASIL library should self-consistently reproduce the far-infrared dust emission. To compare with this, we included in addition a far-infrared 
dust component to explicitly match the starburst properties to the sub-millimetre observations of radio galaxies and quasars by Archibald et al.~(2001), Willott et al.~(2002) and Reuland et al.~(2004). 
From Willott et al.~(2002), the radio quasars appear to be a factor of several brighter than the radio galaxies at 850\micron. In Grimes, Willott \& Rawlings~(2005), this was enshrined as a universal 
relationship between L([OIII]$\lambda5007$) [W] and L(850\micron) [W/Hz/sr]:

\begin{equation}
\rm{log L(850\micron)} = -2.5753 + 0.682\rm{log L([OIII]\lambda5007}),
\end{equation}

with additional scatter drawn from a gaussian of $\sigma = 0.44$. With this normalisation, we adopt a modified blackbody with $\beta = 1.95$, $T=41$\K, as found by Priddey \& McMahon~(2001) for high-redshift quasars. Archibald et al.~(2001) found L(850\micron~(rest)) to be a strongly increasing function of redshift, but for our adopted SED parameters, the redshift dependence is a more modest $(1+z)^{1.5}$ (see Reuland et al., Fig.~4); this dependence is added to eqn.~(8) and normalised at $z=2.5$. There is evidence for a decline in L(850\micron~(rest)) at $z>4$, so in common with other parts of the simulation we assume a decline of the form $(1+z)^{-7.9}$ at $z>4.8$. We note, however, that Rawlings et al.~(2004) found any explicit redshift dependence to be small or absent. Although Willott et al.~(2002) found tentative evidence for an anti-correlation between radio-source size and sub-mm luminosity for quasars, no such trend was found for the radio galaxy sample of Reuland et al.~(2004) and we have chosen not to incorporate one. 

Finally, we remark that our treatment of the infrared SED is confined to reprocessed dust emission due to star formation and AGN heating. As a result, we have not endeavoured to incorporate an infrared synchrotron emission component in the beamed radio-loud AGN. If required, such a component could be added as an extrapolation of the radio SED given in the $S^{3}$ database, following the beaming treatment described by W08. It would only have a significant effect on the predicted source counts at the brightest flux densities.

\section{COMPARISON WITH EXISTING OBSERVATIONS}
\begin{figure*}
\includegraphics[width=0.47\textwidth,angle=0]{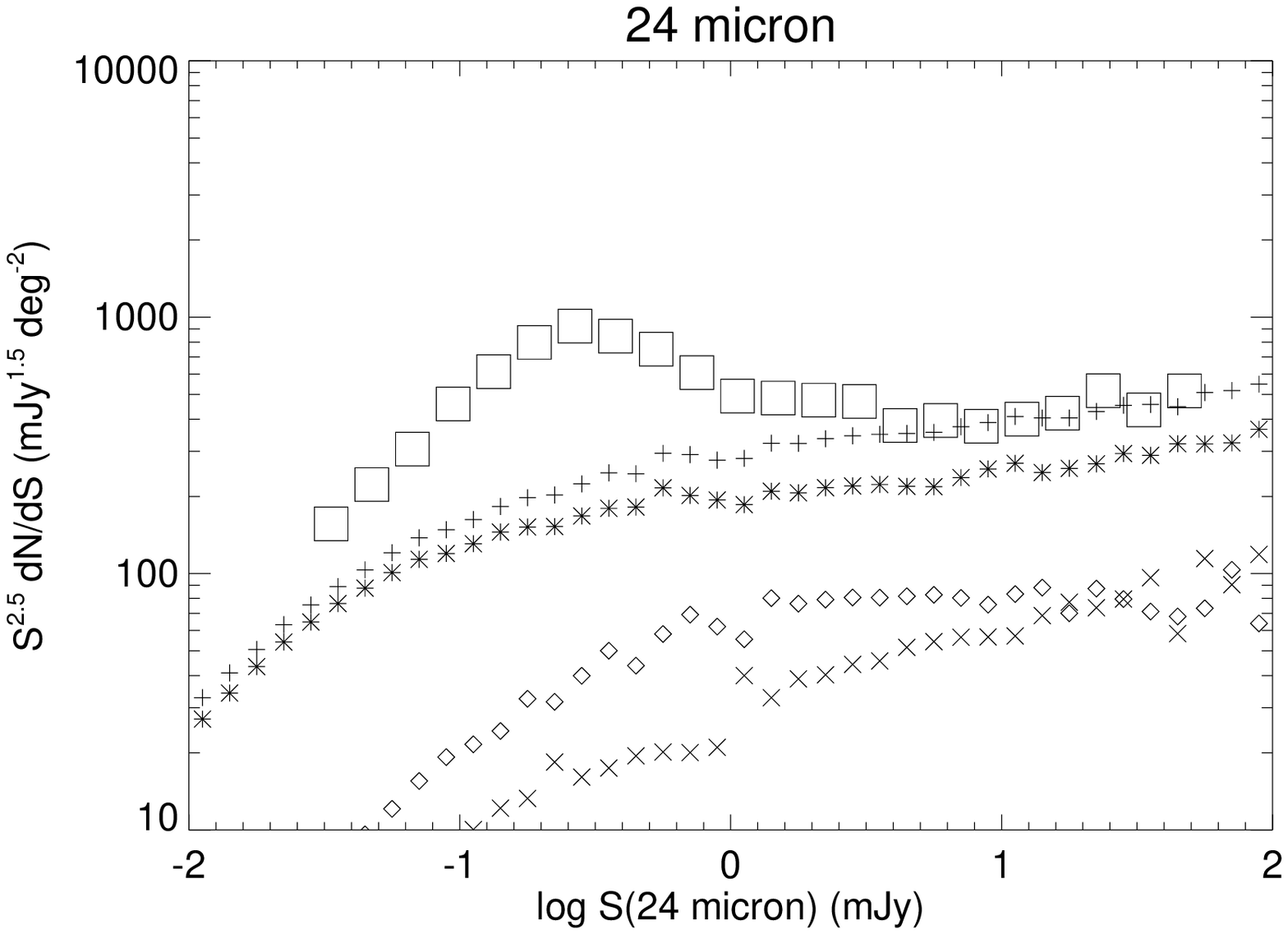}
\includegraphics[width=0.47\textwidth,angle=0]{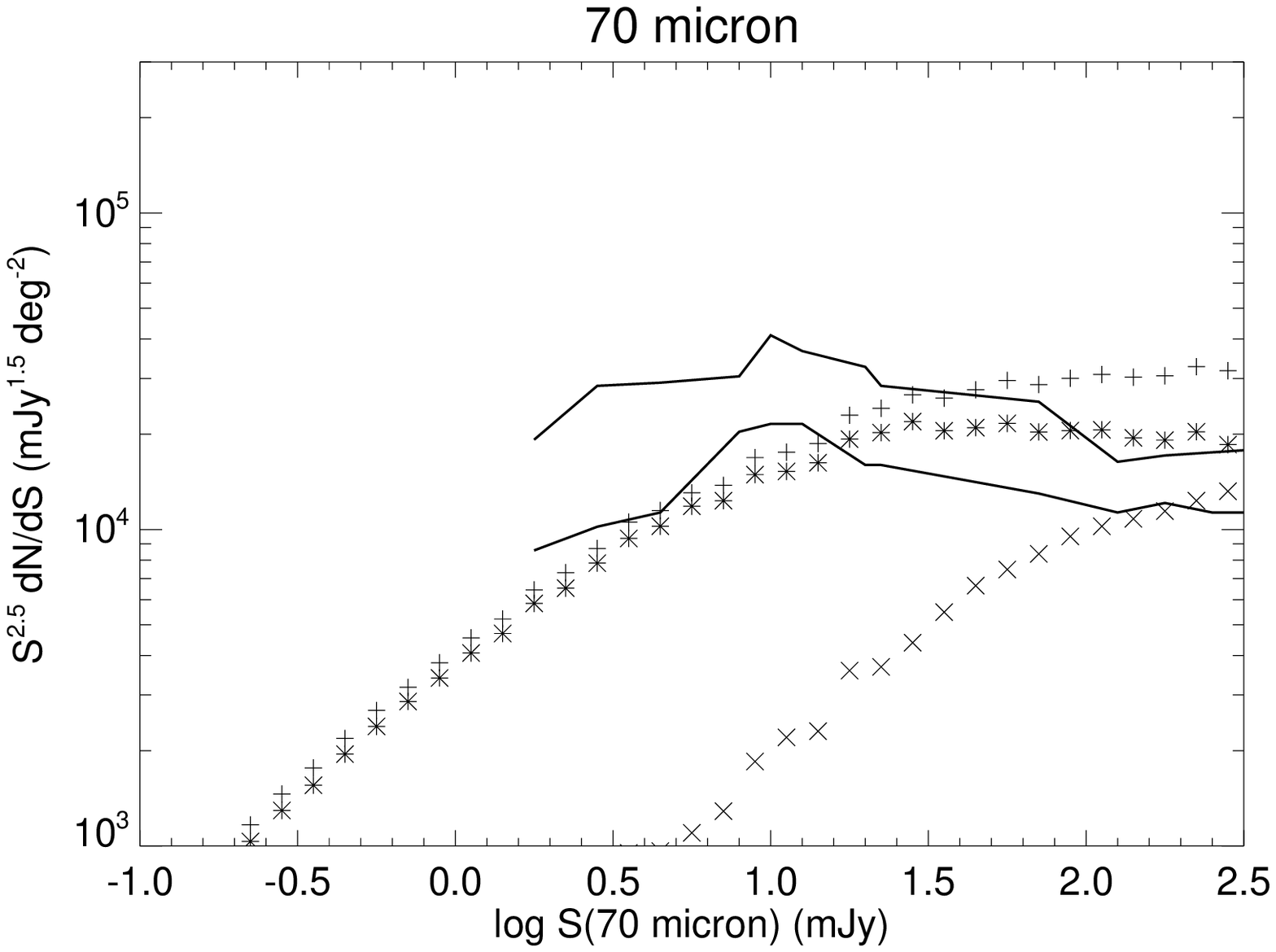}
\includegraphics[width=0.47\textwidth,angle=0]{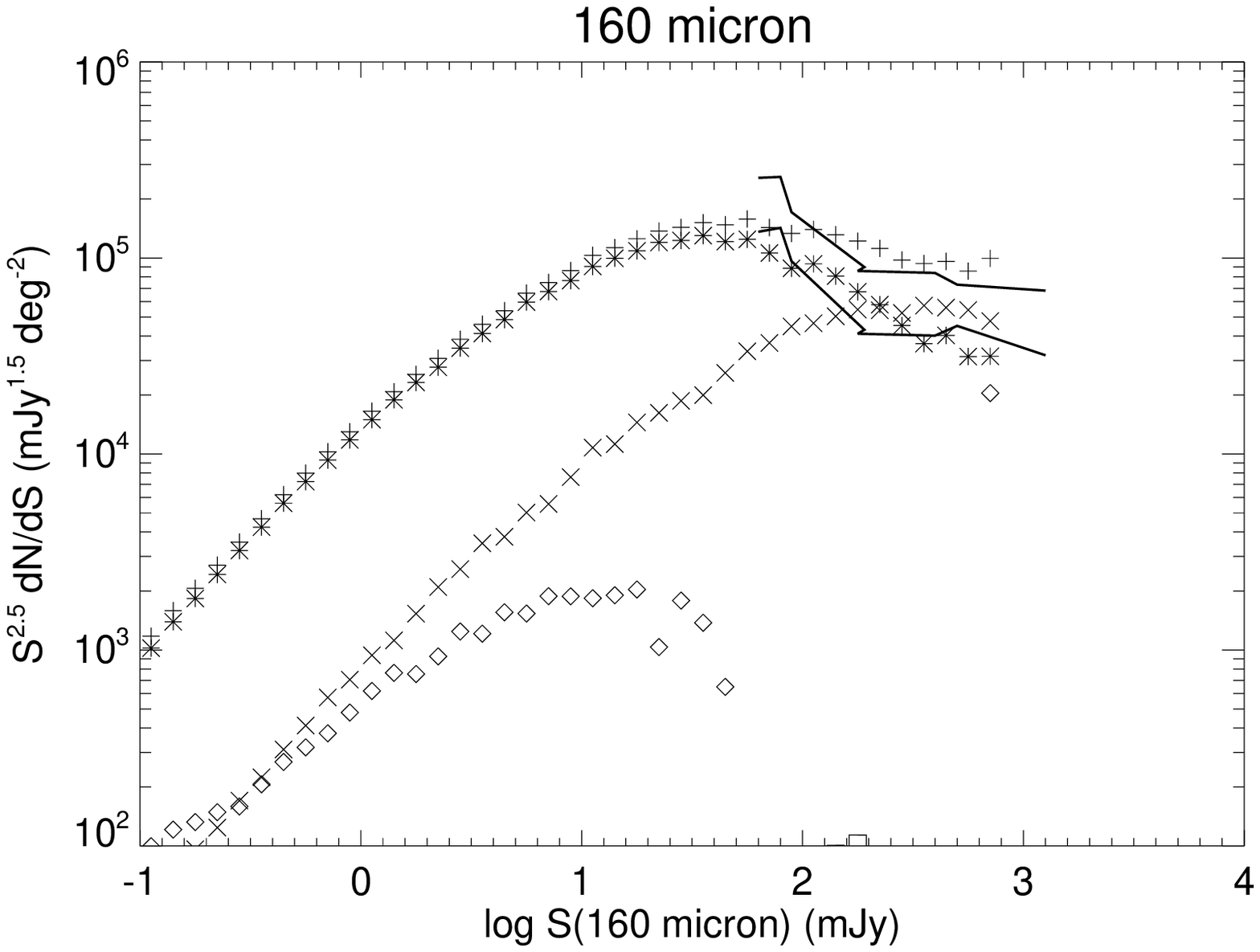}
\caption{\normalsize Source counts for the baseline model simulation (section 2) showing normal (asterisks) and starburst galaxies (cross), radio-quiet AGN (diamonds), and the total (plus signs) [FRIs and FRIIs are also included in the totals]. The large squares on the 24\micron~panel show the {\em Spitzer} counts of Papovich et al.~(2004); the thick lines in the 70 and 160\micron~panels bound the observed counts in the S-COSMOS and SWIRE fields (Frayer et al.~2009). The plots were generated using a $4$~deg$^{2}$ region of the W08 input radio simulation and the full $400$~deg$^{2}$ to generate the bright flux end (see section 2 for details).}
\label{fig:BASELINEdsc}
\end{figure*}

\begin{figure}
\includegraphics[width=0.47\textwidth,angle=0]{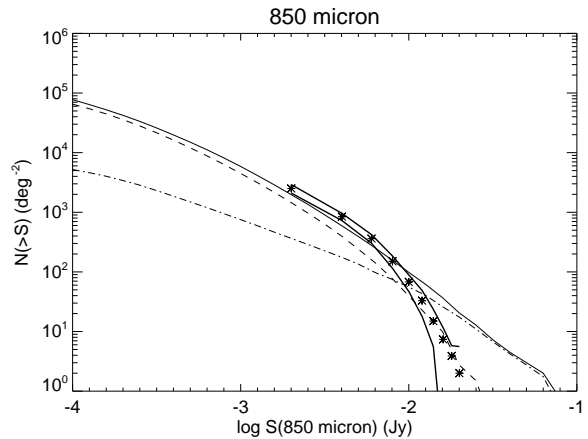}
\caption{\normalsize The thin continuous line shows the simulated integral source counts at 850\micron~for the baseline model using a 4~deg$^{2}$ region of the input radio simulation; the contributions of normal and starburst galaxies are shown by the dashed and dot-dashed lines, respectively. The asterisks and thick lines show the measured counts and uncertainty region for the SCUBA Half-Degree Extragalactic Survey (Coppin et al. 2006).}
\label{fig:BASELINE850csc}
\end{figure}

The prescriptions outlined in section 2 constitute our `baseline model'. Here we compare its source counts and other derived quantities against observations. In Fig.~\ref{fig:BASELINEdsc} we show the normalised differential source counts at 24, 70 and 160\micron~for comparison with the most recent {\em Spitzer} MIPS survey results. To generate these plots we used a 4~deg$^{2}$ area of the input radio simulation, coupled with a shallow subset of the full 400~deg$^{2}$ radio simulation area to generate the bright flux ends (above 1.6\mJy~at 24\micron, 30\mJy~at 70\micron~and 100\mJy~at 160\micron). The simulated counts fall decisively short of the observations below 1\mJy~and 10\mJy~at 24 and 70\micron, respectively. There is a modest excess of simulated sources above $\sim 100$\mJy~at 70 and 160\micron, and above $\sim 10$\mJy~at 850\micron~(Fig.~\ref{fig:BASELINE850csc}) where we compare with results from the {\em SCUBA} Half Degree Extragalactic Survey (SHADES; Coppin et al.~2006).

\begin{figure}
\includegraphics[width=0.47\textwidth,angle=0]{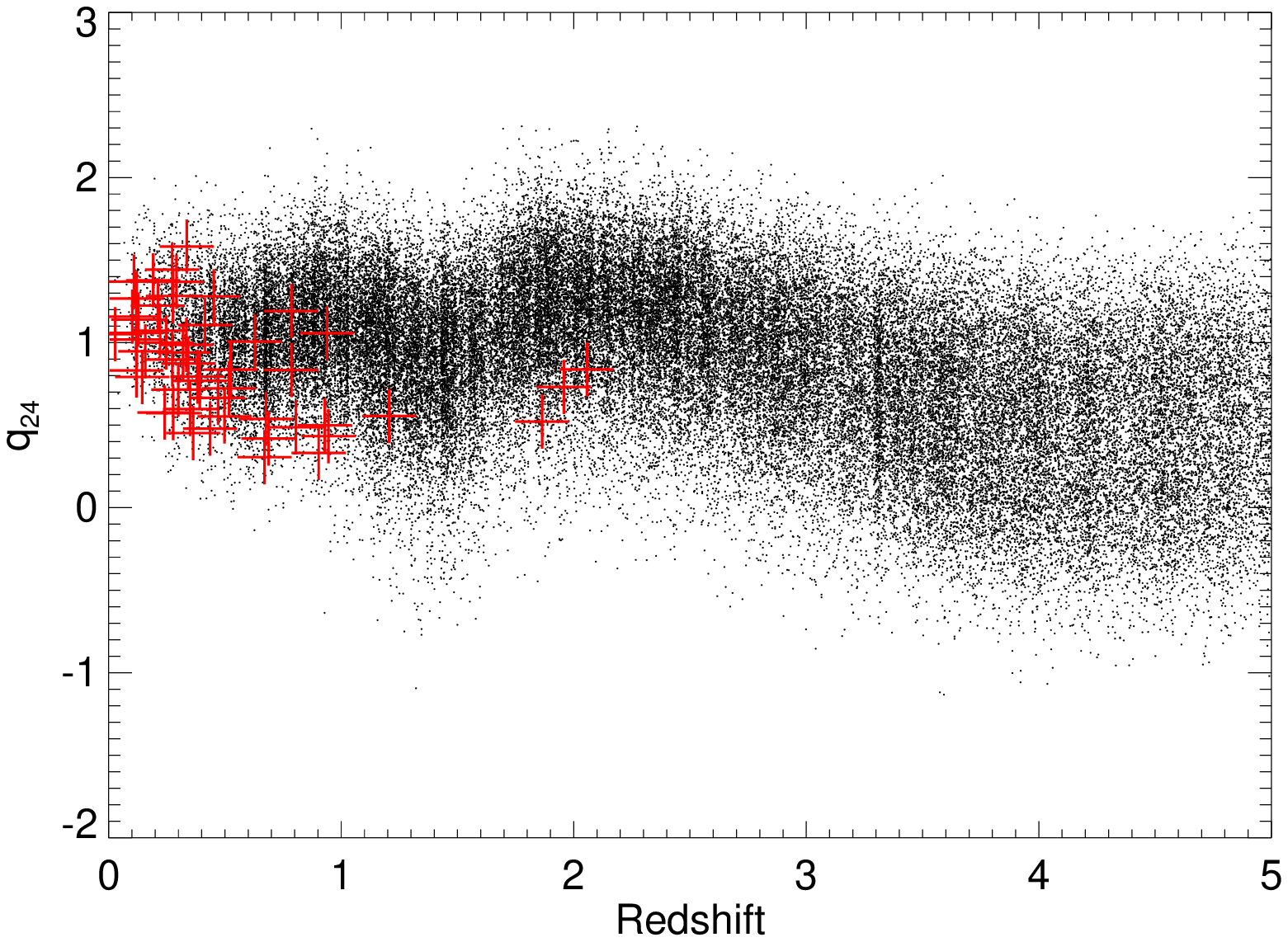}
\includegraphics[width=0.47\textwidth,angle=0]{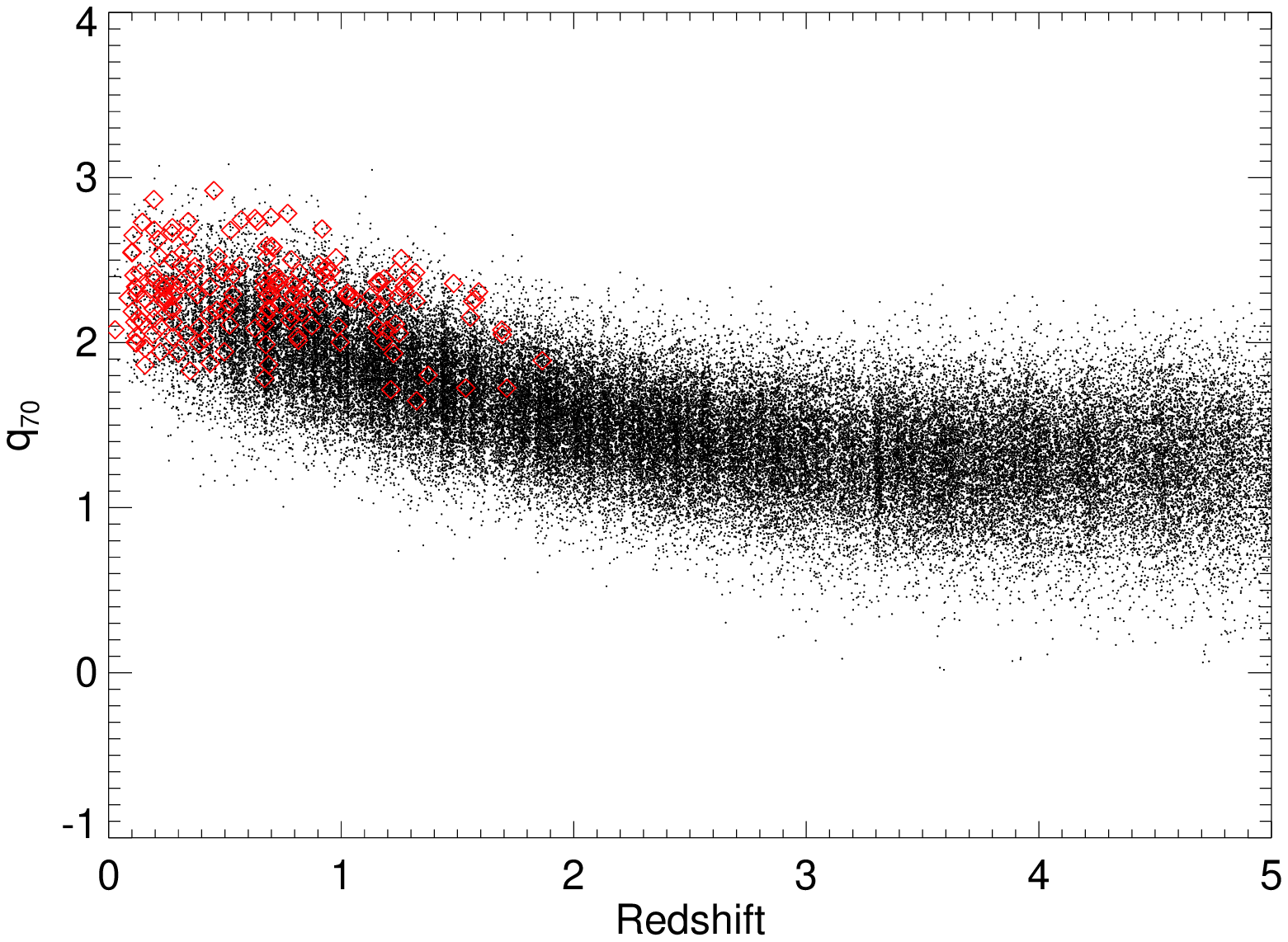}
\caption{\normalsize The quantity $q_{\rm{IR}} = log (F_{\rm{IR}}/F_{\rm{20 cm}})$ is used to characterise the far-infrared:radio correlation for star-forming galaxies and is plotted here for the entire population of simulated normal galaxies (black dots) at 24 and 70\micron, using the baseline model of section 2. The red crosses and diamonds are the sources remaining after the application of infrared and radio flux cuts to mimic the selection of Appleton et al.~(2004) and Seymour et al.~(2009), respectively. }
\label{fig:BASELINEqparams}
\end{figure}

Since the source counts are dominated by star-forming galaxies, we show in Fig.~\ref{fig:BASELINEqparams} the quantity $q_{\rm{IR}} = log (F_{\rm{IR}}/F_{\rm{20 cm}})$ in the 24 and 70\micron~bands for the simulated normal galaxy population. $F_{\rm{IR}}$ and $F_{\rm{20 cm}}$ are the simulated infrared and radio fluxes which would be observed in these bands, i.e. without the application of any K-corrections. The simulated sample is dominated by galaxies substantially fainter than those in current observational samples, but our $q_{\rm{24}}$ and  $q_{\rm{70}}$ values compare reasonably well with those measured locally (Appleton et al.~2004, Beswick et al.~2008 and Garn et al.~2009a,b). For a better comparison with the observations, we applied cuts in flux density to yield the sub-samples also shown in Fig.~\ref{fig:BASELINEqparams}: at 24\micron, we selected galaxies with $F_{\rm{24}} > 0.3$\mJy~and $F_{\rm{20 cm}} > 90$\microJy~to mimic the selection of Appleton et al.~(2004), but we note that these authors only included sources with spectroscopic redshifts and appear not to discriminate between star-forming galaxies and AGN, biases which we cannot easily accommodate. At 70\micron, we apply cuts of $F_{\rm{70}} > 6$\mJy~and $F_{\rm{20 cm}} > 30$\microJy~to match the selection of Seymour et al.~(2009), who use photometric and spectroscopic redshifts and apply star-forming galaxy/AGN classification. The resulting sample shows only a very modest decline in $q_{\rm{70}}$ out to $z=2$, consistent with the findings of Seymour et al.

\section{REVISIONS TO THE BASELINE MODEL}
We now adapt the baseline model to provide a better match to the existing infrared source count data. We focus entirely on the star-forming galaxies and perform the modifications in two steps, motivated by recent 
observational findings. Firstly, we use post-processing to modify the cosmological luminosity evolution of the star-forming galaxies to remove the shortfall in the 70\micron~source counts. Secondly, we allow the SED template 
assignment procedure to evolve towards cooler templates at higher redshift in an attempt to rectify the deficit in the 24\micron~counts.

\subsection{Modifications to the luminosity evolution of the star-forming galaxies} 
\begin{figure}
\includegraphics[width=0.47\textwidth,angle=0]{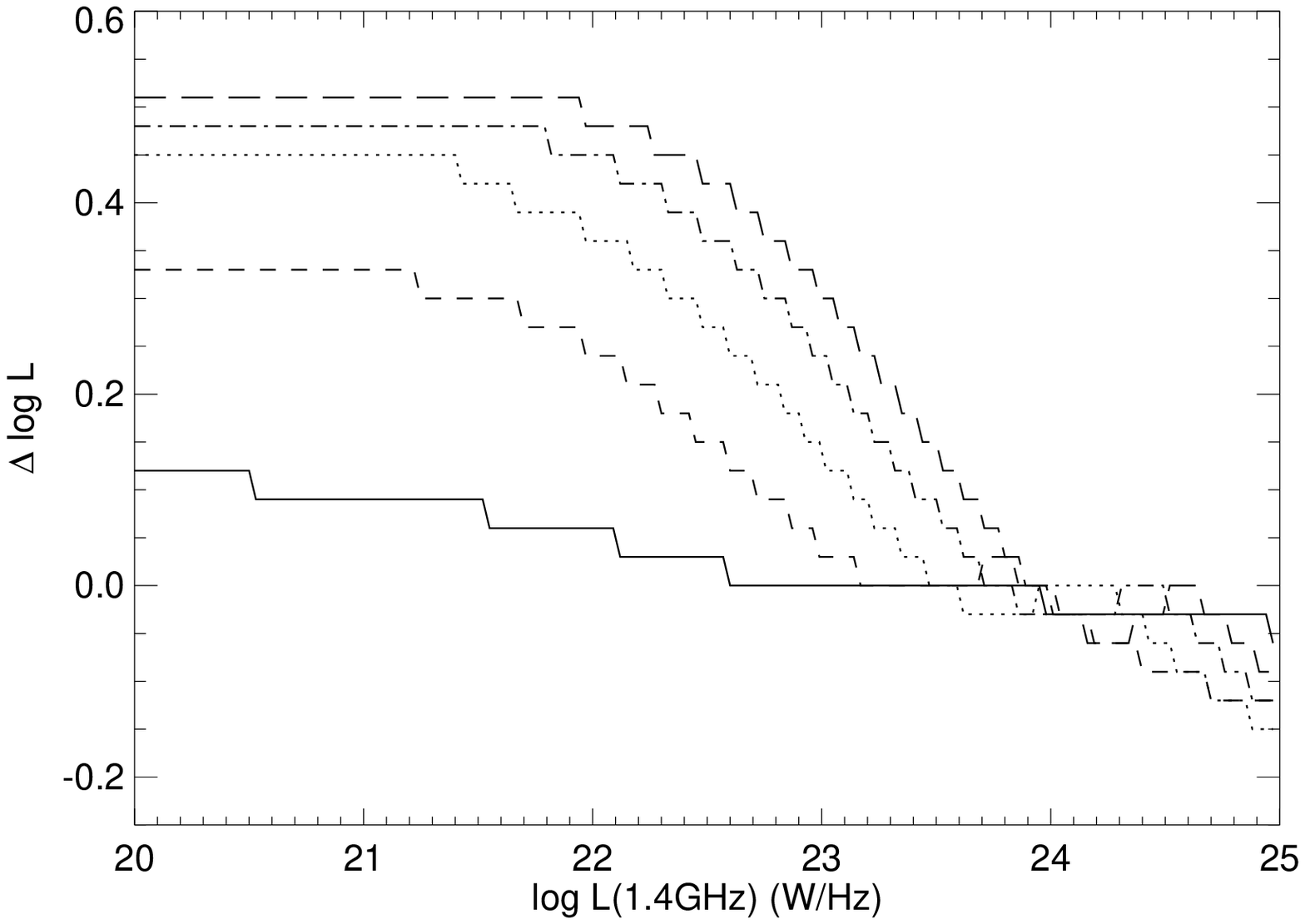}
\caption{\normalsize The boost in luminosity, $\Delta log L(L,z)$ (dex), which must be applied to the star-forming galaxies in the W08 catalogue to transform to pure luminosity evolution of the form $(1+z)^{3.1}$ to $z=1.5$ in a
$\Lambda$-Universe. The lines show redshifts z=0.08 (solid), 0.45 (short dashed), 0.83 (dotted), 1.2 (dot-dashed) and 1.5 (long dashed).} 
\label{fig:lumboost}
\end{figure}

\begin{figure*}
\includegraphics[width=0.47\textwidth,angle=0]{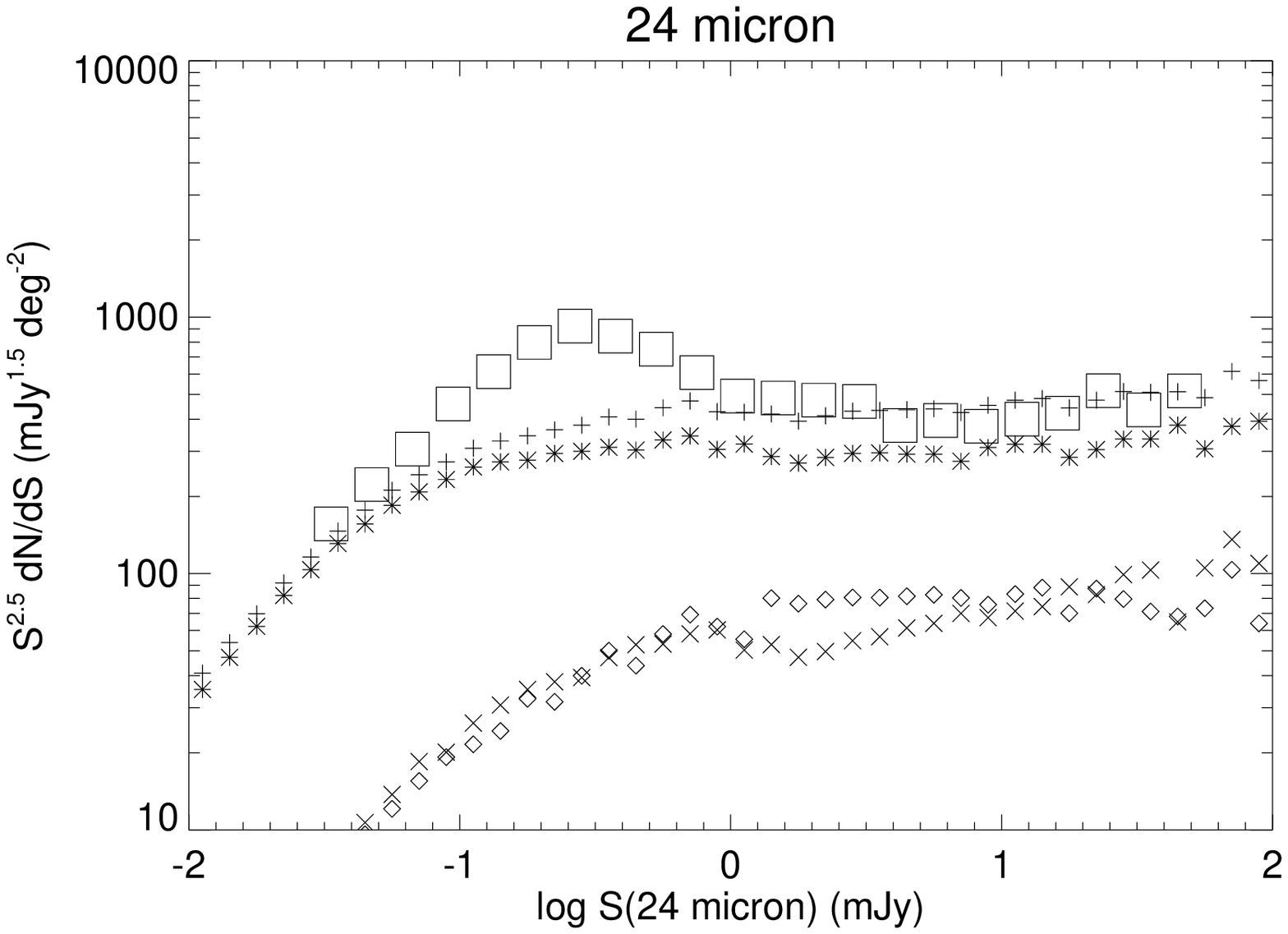}
\includegraphics[width=0.47\textwidth,angle=0]{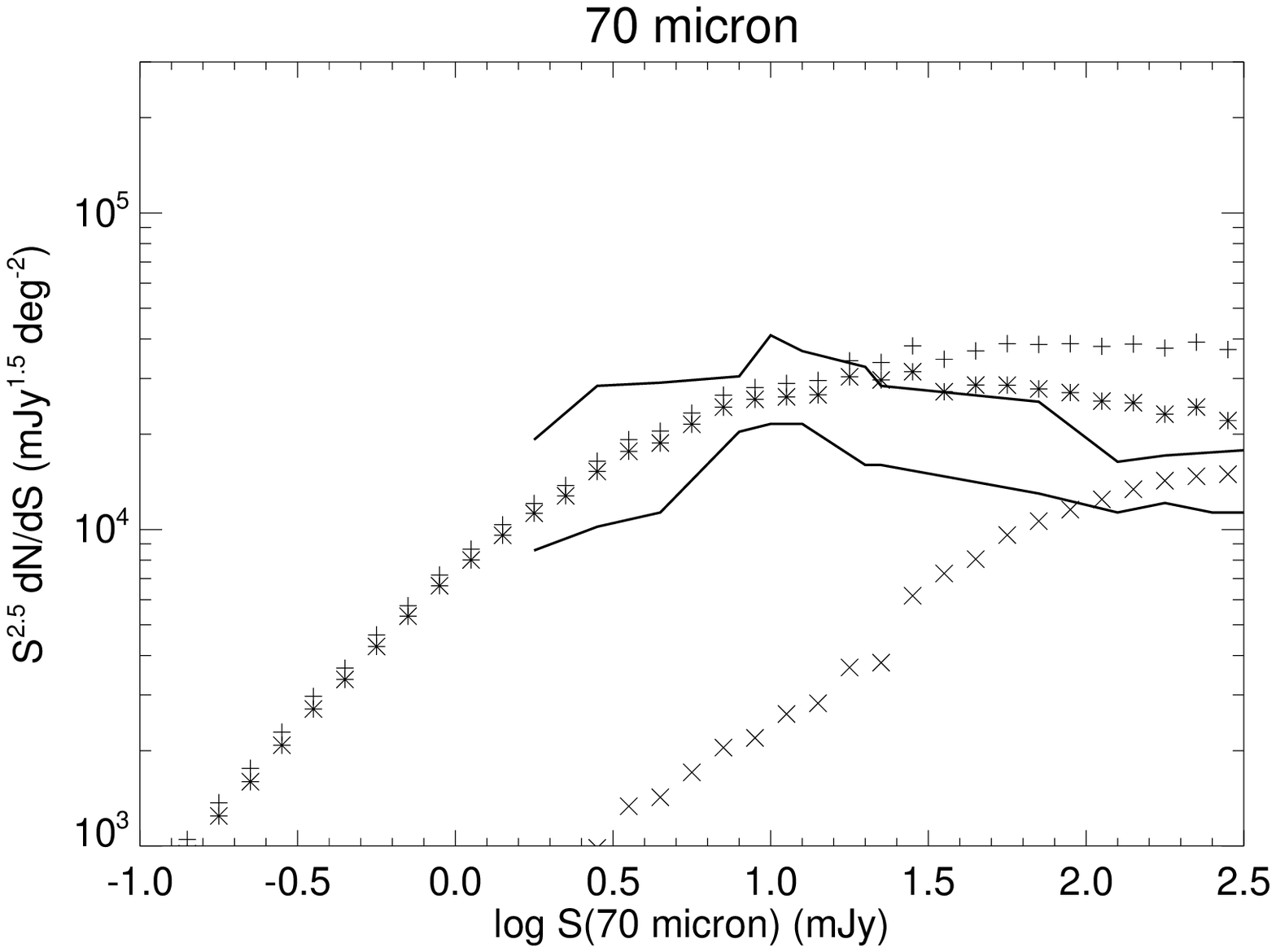}
\includegraphics[width=0.47\textwidth,angle=0]{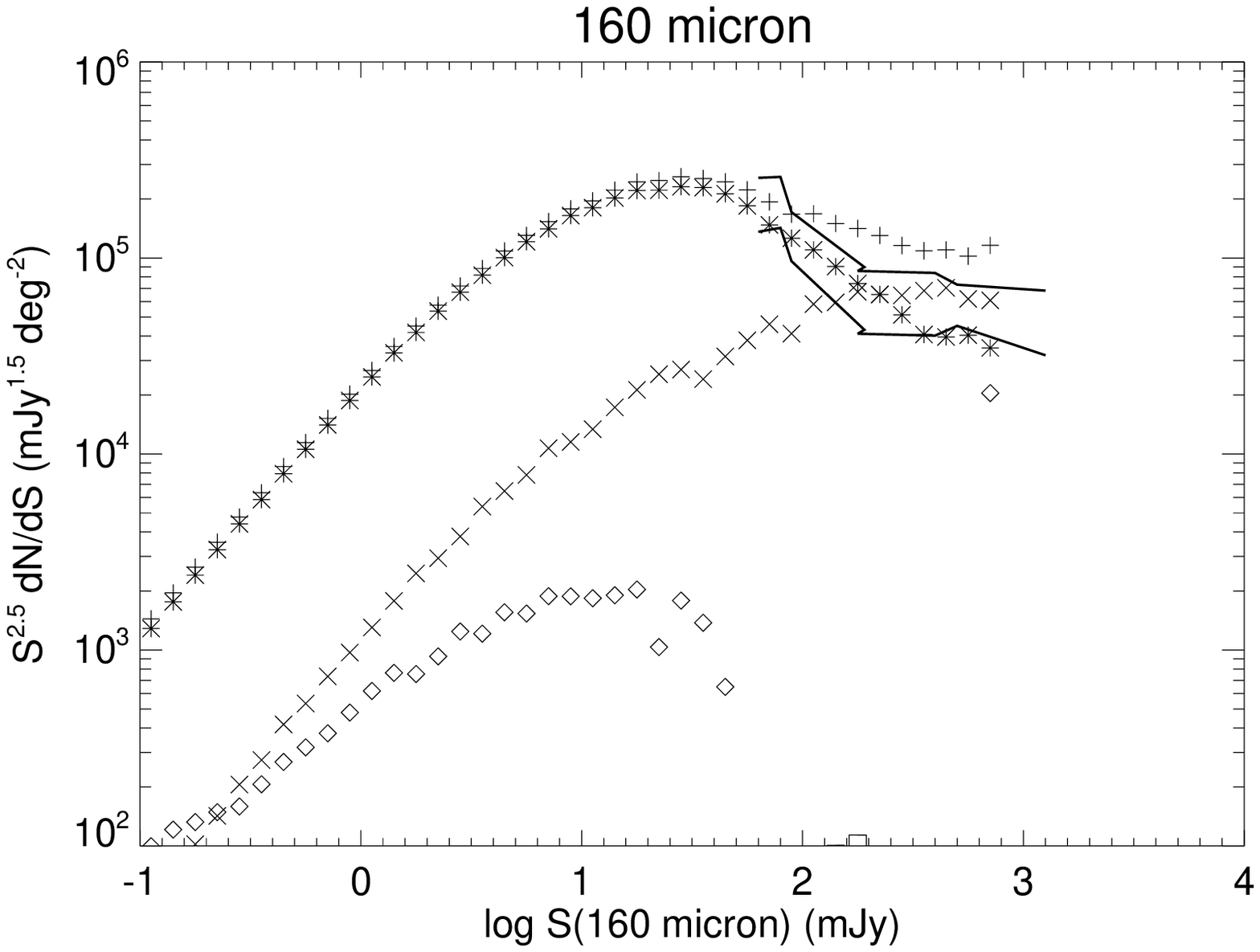}
\caption{\normalsize  24, 70 and 160\micron~source counts for the modified model of section 4.1, in which the luminosity of the star-forming galaxies is boosted to mimic pure luminosity evolution in a $\Lambda$-universe. 
Symbols and observed data are as in Fig.~\ref{fig:BASELINEdsc}.}
\label{fig:LUMBOOSTdsc}
\end{figure*}

In W08 we assumed that the radio luminosity function of the star-forming galaxy population undergoes pure luminosity evolution (PLE) of the form $(1+z)^{p}$ with p=3.1 out to z=1.5, 
applied in an {\em Einstein -- de Sitter cosmology}, but adapted to the flat $\Lambda$ cosmology of the simulation. Since then, new observational constraints on the evolutionary form have been published. At 
70\micron, Huynh et al.~(2007) constrained the evolution out to $z \sim 1$ and found a degeneracy between luminosity and density evolution with $p=2.78$ in the PLE case; using several {\em Spitzer} surveys, 
Magnelli et al.~(2009) found redshift evolution consistent with PLE with $p=3.6 \pm 0.4$ out to $z \sim 1.3$. Both these studies derived the evolution for a flat $\Lambda$ cosmology, which for a given functional form of PLE, leads to a higher 
abundance of galaxies than present in the W08 simulation. The discrepancy is especially pronounced at and below the break in the luminosity function. We can, however, correct for it by boosting the luminosity 
of the star-forming galaxies in the W08 catalogue by an amount $\Delta log L(L,z)$. The boost required to bring the W08 luminosity function into agreement with PLE of $p=3.1$ to $z=1.5$ in a $\Lambda$ universe is shown 
in Fig.~\ref{fig:lumboost}. [We note, however, that W08 assumed that the local luminosity function is flat below 20.6 W/Hz, in which regime no luminosity boost can compensate for the deficit in space density; for continuity, 
the boost is accordingly held constant below this evolving break luminosity]. After the application of this luminosity boost, the 24 and 70\micron~source counts are as shown in Fig.~\ref{fig:LUMBOOSTdsc}. The 70 \micron~counts below 10\mJy~are now in satisfactory agreement with the observational data but a deficit remains at 24\micron, which we address in the next subsection.

\subsection{Evolution towards cooler dust templates and the final model}

There is growing evidence that the one-to-one correspondence between infrared luminosity and template SED established in the local Universe breaks down at earlier cosmic epochs. To cite several 
examples, Symeonidis et al.~(2009) showed that luminous and ultraluminous galaxies out to redshift $z \sim 1$ have much cooler dust distributions (i.e. SEDs peaking at longer wavelengths) 
than their local counterparts in the IRAS Bright Galaxy Sample (see also Symeonidis et al.~2008); and in the mid-infrared, Farrah et al.~(2008) showed that ULIRGs at $z \sim 1.7$ have 
spectral features characteristic of starbursts with luminosities $10^{10-11}$\Lsun, rather than local ULIRGs. Such trends may reflect more spatially extended star-forming regions at 
higher redshifts with lower dust optical depths. Based on a study of sub-mm galaxies at $z \sim 2.5$, Chapin et al.~(2009) argued for luminosity evolution in the colour-luminosity relation 
of the form $(1+z)^{3}$, in the sense that the SED becomes cooler at higher redshift (see also Pope et al.~2006).

\begin{figure}
\includegraphics[width=0.47\textwidth,angle=0]{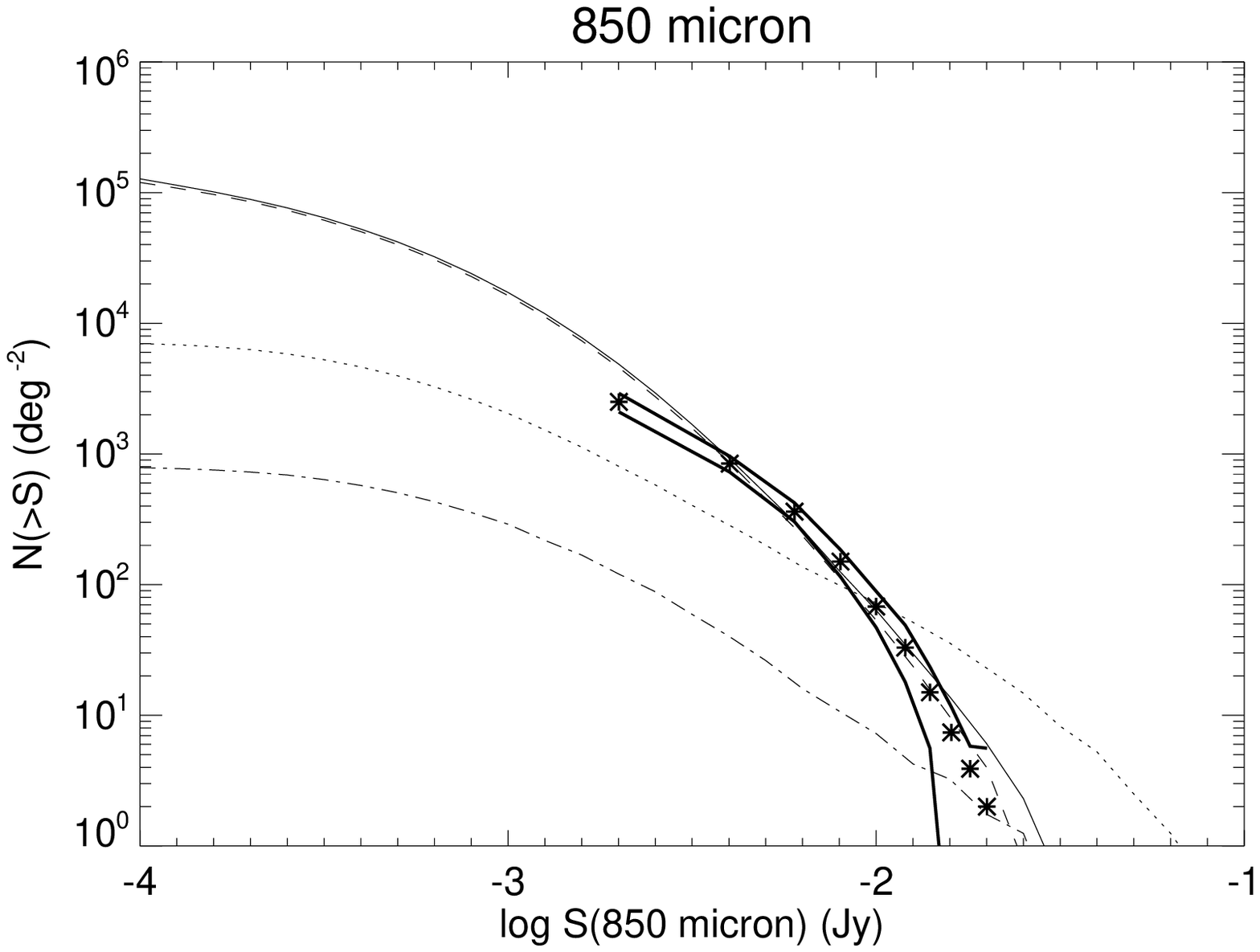}
\includegraphics[width=0.47\textwidth,angle=0]{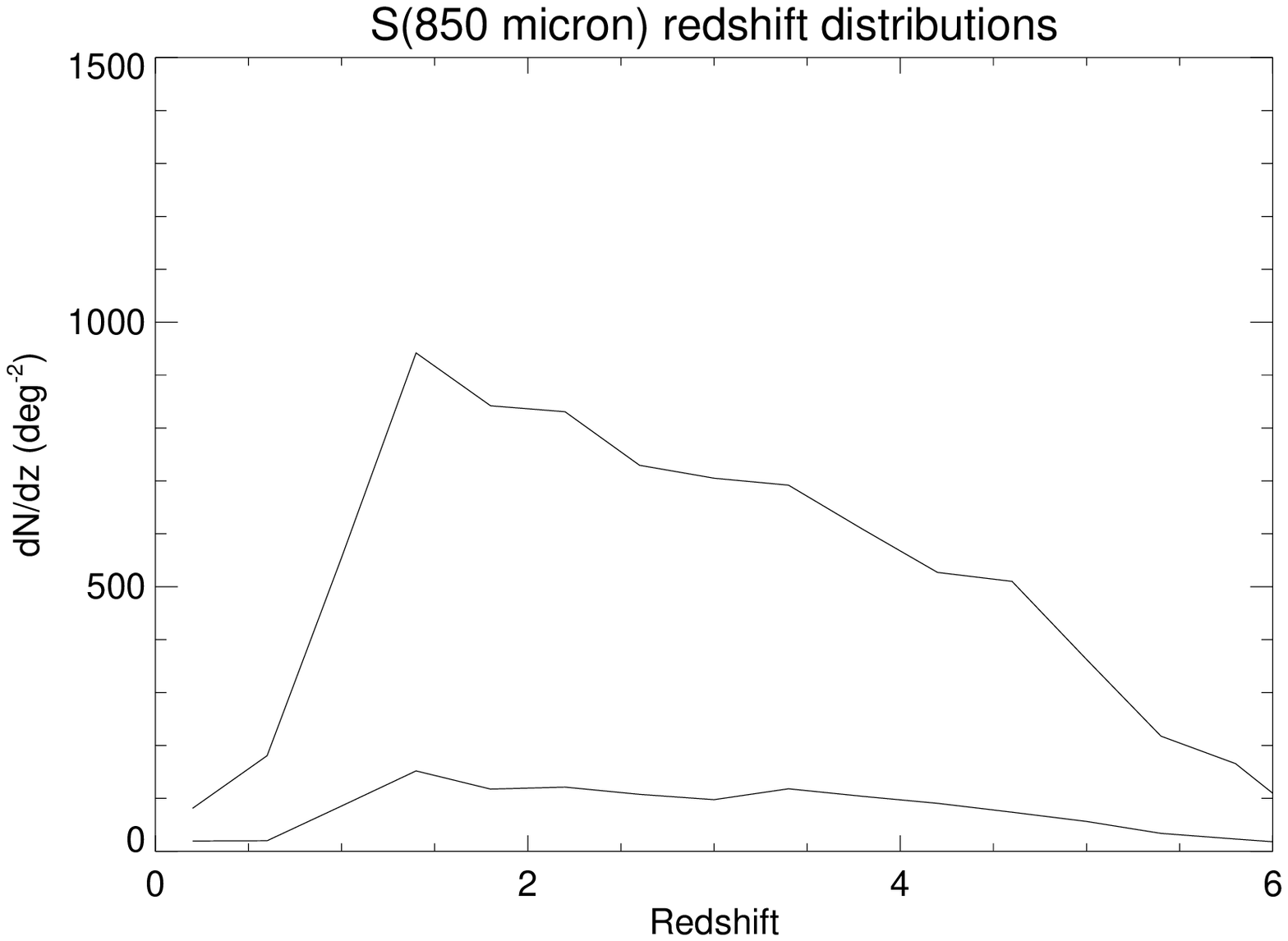}
\caption{\normalsize Upper panel: 850\micron~source counts for the final model of section 4.2 as determined from a 4~deg$^{2}$~simulation area; the observed SHADES counts are the same as in Fig.~\ref{fig:BASELINE850csc}; the dashed line shows the contribution of the normal galaxies and the dot-dashed line represents the starbursts, assuming a sharp cut-off in the latter population at $z>1.5$; the dotted line shows the contribution of the starbursts under the assumption of the `default post-processing' cut-off in the space density, i.e. $(1+z)^{-7.9}$ at $z>4.8$. Lower panel: 850\micron~redshift distributions for flux limits of 2.4 and 5\mJy, assuming the $z>1.5$ cut-off in the starburst population.}
\label{fig:TEMPEVdsc850}
\end{figure}

Building on these findings, we now introduce phenomenological modifications to the SED template assignment procedure with the aim of reducing the remaining deficit in the 24\micron~source counts 
(Fig.~\ref{fig:LUMBOOSTdsc}). With reference to the Rieke templates in Fig.~\ref{fig:RiekesameFIR}, we see that for a given L(FIR), evolution towards cooler dust templates with increasing redshift 
can significantly boost the observed 24\micron~flux relative to that at 70\micron. Accordingly, the L(FIR) is scaled down by a factor $f(z)$ for the purposes of selecting the far-infrared colour, $C$,
from the $C$--L(FIR) distribution (see section 2.1). The chosen template is, however, still normalised to the original value of L(FIR). After some experimentation, we chose a function which evolves as 
$f(z) = (1 + z)^{2.5}$ out to $z=1$. Beyond $z \sim 1$, the evolution towards cooler templates must stop or go into reverse in order not to overproduce the 850\micron~counts. We chose a decline of
the form $(1+z)^{-1.5}$ from $z=1-2$, flat thereafter. Even with this in place, the starbursts still overproduce the bright end of the 850\micron~source counts above 10\mJy~if we retain the `default
post-processing option', i.e. a $(1+z)^{-7.9}$ decline in space density at $z>4.8$. As shown in Fig.~\ref{fig:TEMPEVdsc850}, one possible way to avoid this is to remove the entire starburst population (i.e. the high-luminosity component) at $z>1.5$ and we adopt this solution for our final model. 

The removal of the high-redshift starburst component of the luminosity function implies that the evolving normal galaxy population in the simulation is sufficient to account for the star formation at these epochs. Although we used the terms `normal galaxy' and `starburst' merely as labels to refer to the two Schechter function components of the local luminosity function, this trend may find some physical justification in the work of Obreschkow \& Rawlings~(2009). The latter authors showed that, with increasing redshift, galaxy disks are smaller, leading to higher gas densities and pressures and hence higher molecular gas fractions and star formation rates. In the local Universe, star formation rates in excess of $\sim 10$\Msunpyr~can only occur in dense gas disks created by zero angular momentum gas in major mergers characteristic of starburst galaxies, e.g. the merger-triggered local ULIRGs. 

Even with the starburst cut-off in place, the modest (factor $\sim 2$) excesses of simulated sources at the bright flux ends of all three {\em Spitzer} bands persist. In Fig.~\ref{fig:fluxredshift}, we show the flux--redshift planes for the normal galaxy and starburst populations for 70 and 160\micron~fluxes exceeding 100\mJy; much of the excess normal galaxy population occurs at redshifts $z<0.5$, particularly at 70\micron. It may arise from a combination of effects, such as a departure from strictly power-law pure luminosity evolution or from an inaccurate treatment of the scatter on the far-infrared relation in eqn.~(1) (i.e. from galaxies scattered into the high-L(FIR) gaussian wing of the ``q'' parameter distribution). In support of the first possibility, Huynh et al.~(2007) found no evidence for significant evolution at infrared luminosities below $10^{11}$\Lsun~for redshifts $z<0.4$.

A modest (factor $\sim 2$) deficit also persists in the 24\micron~counts around 0.3\mJy, which suggests that the required template evolution may be more extreme than assumed, possibly requiring a dependence on luminosity in addition to redshift. It is also possible that the Rieke templates underestimate the PAH strength in the higher redshift galaxies.

Finally, we note that gravitational lensing biases (e.g. Negrello et al.~2007) are not accounted for in these simulations, but can be critical when the source counts are very steep.

\begin{figure*}
\includegraphics[width=0.47\textwidth,angle=0]{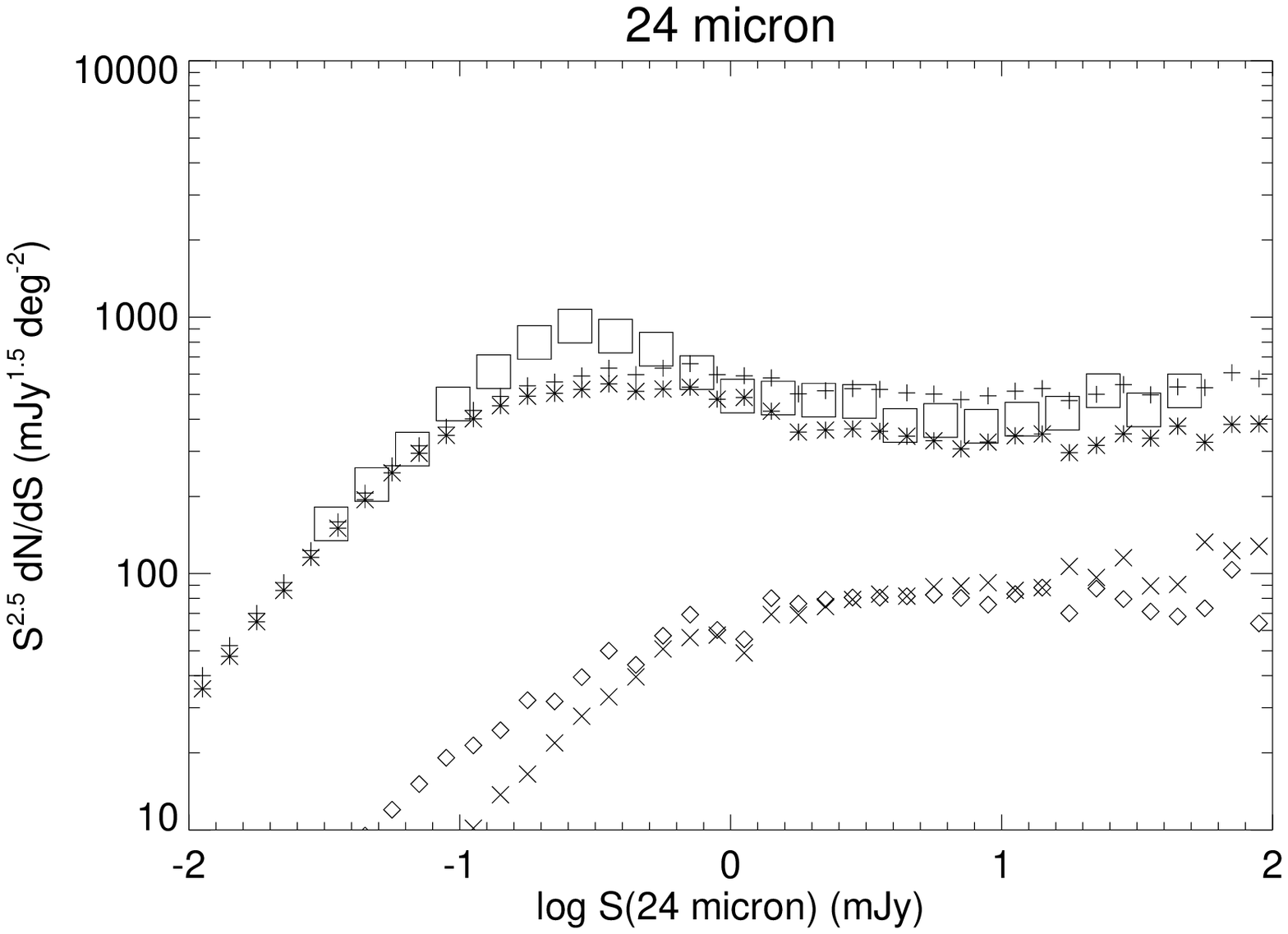}
\includegraphics[width=0.47\textwidth,angle=0]{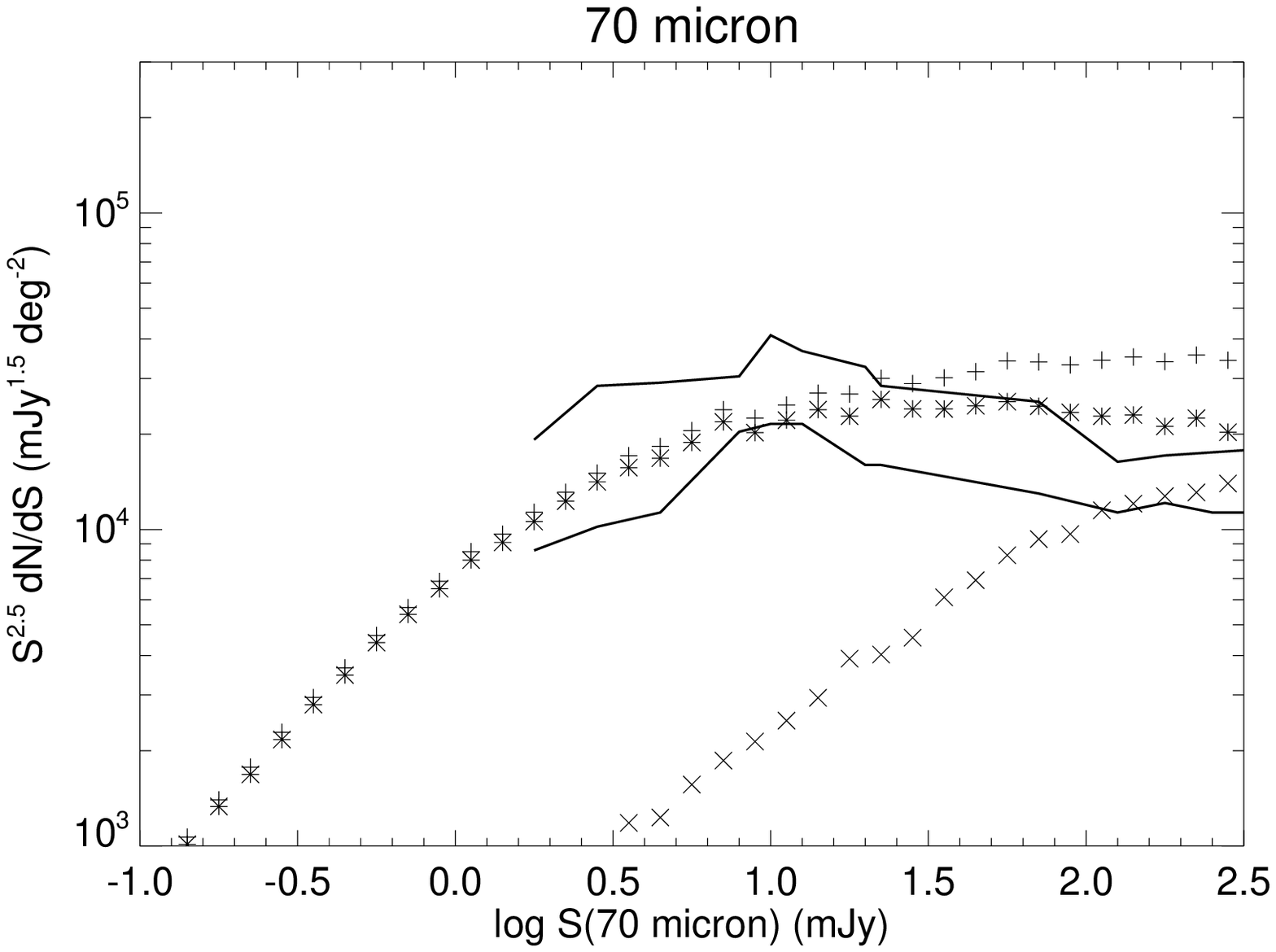}
\includegraphics[width=0.47\textwidth,angle=0]{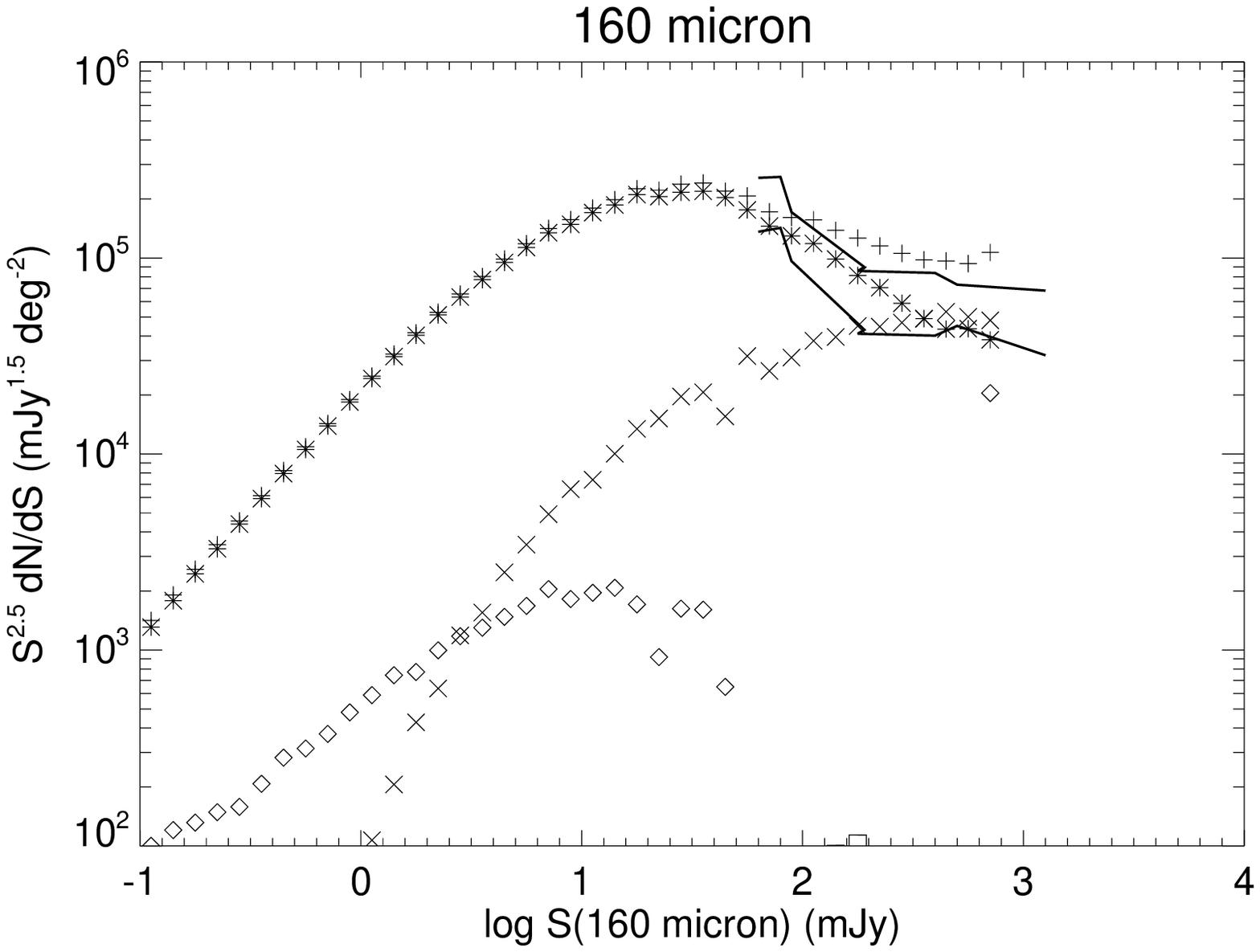}
\caption{\normalsize Spitzer MIPS source counts for the final model, after application of a luminosity boost (section 4.1) and template evolution (section 4.2) to 
the star-forming galaxies and with the starburst population removed at $z>1.5$; symbols and observed data are as in Fig.~\ref{fig:BASELINEdsc}.}
\label{fig:TEMPEVdsc}
\end{figure*}

\begin{figure}
\includegraphics[width=0.47\textwidth,angle=0]{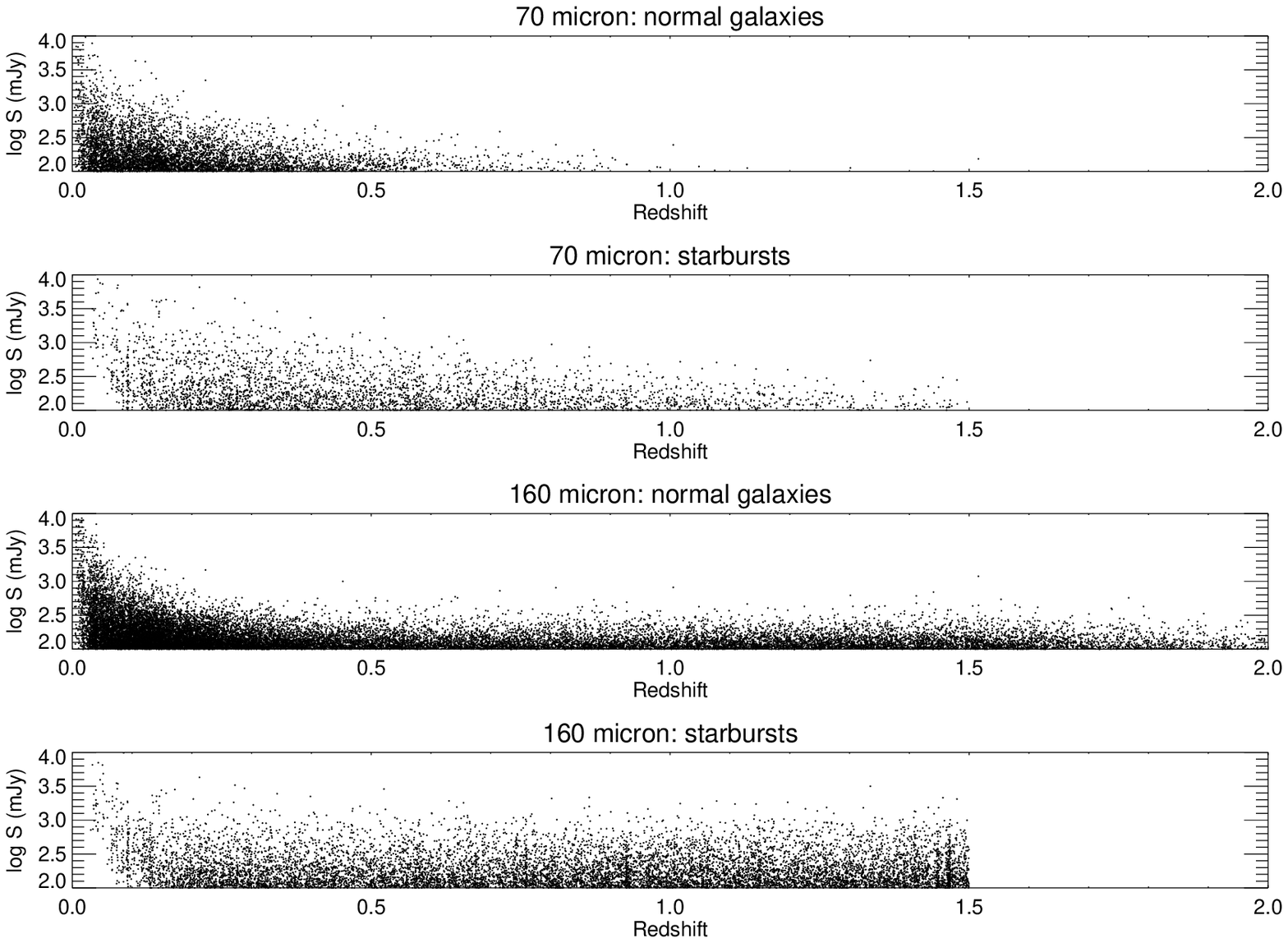}
\caption{\normalsize Flux-redshift distributions for simulated normal galaxies and starbursts contributing to the excess differential source counts above 100\mJy~in Fig.~\ref{fig:TEMPEVdsc} at 70 and 160\micron, demonstrating that much of this excess arises from a local population at $z<0.5$; determined using a shallow subset of the full $20 \times 20$~deg$^{2}$ input radio simulation.}
\label{fig:fluxredshift}
\end{figure}

\subsection{Comparison with observed redshift distributions}
\begin{figure*}
\includegraphics[width=0.47\textwidth,angle=0]{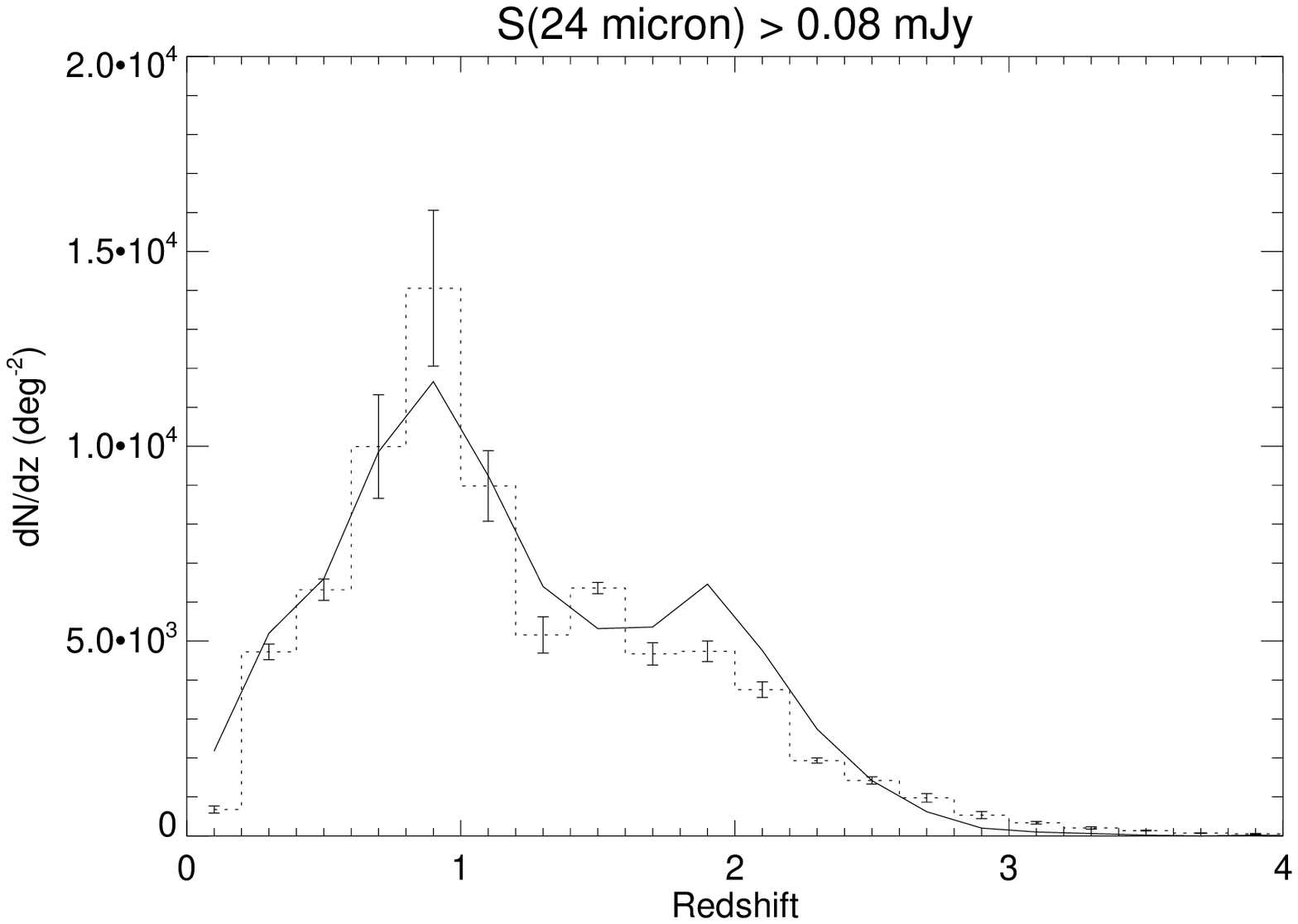}
\includegraphics[width=0.47\textwidth,angle=0]{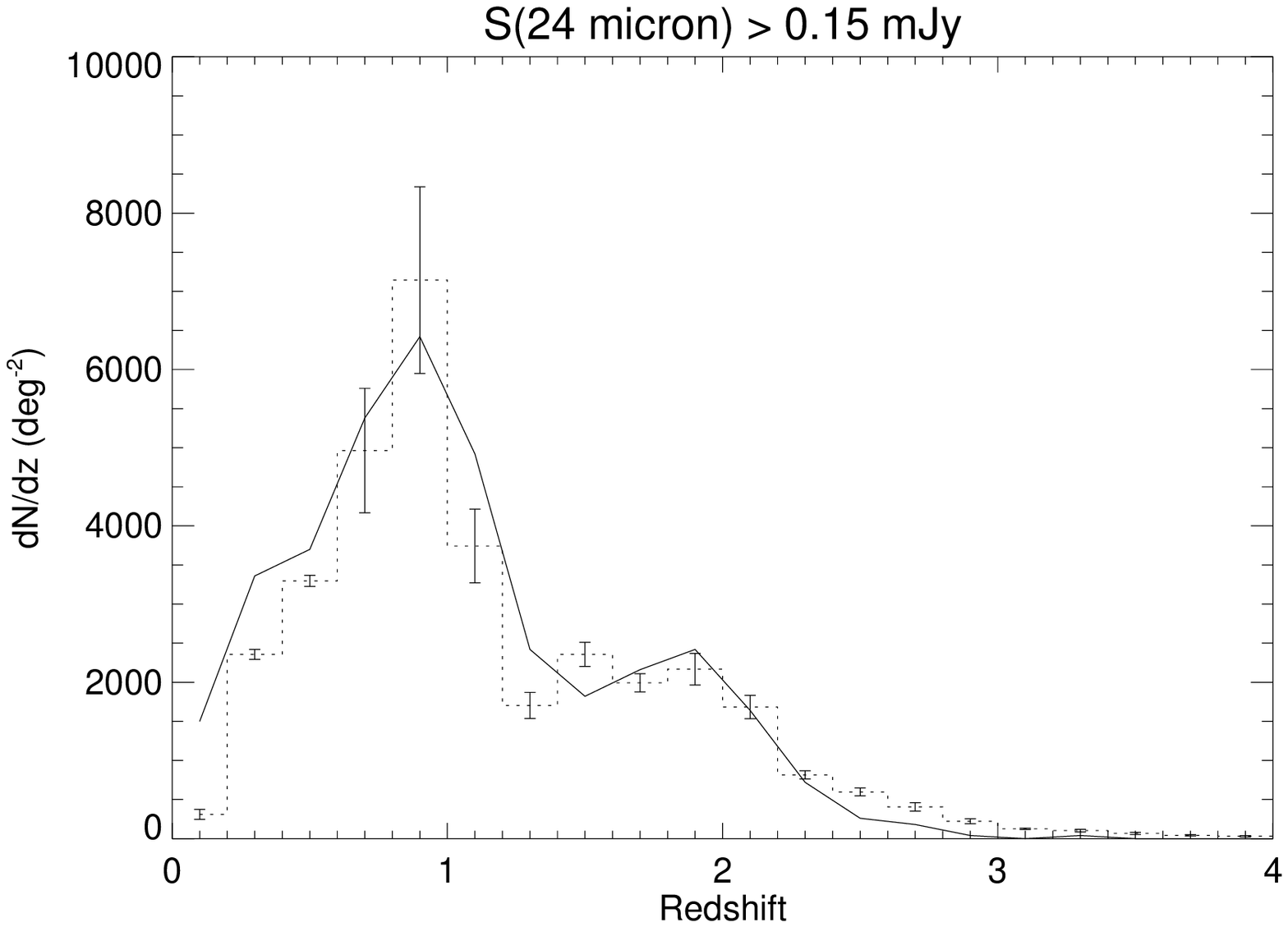}
\includegraphics[width=0.47\textwidth,angle=0]{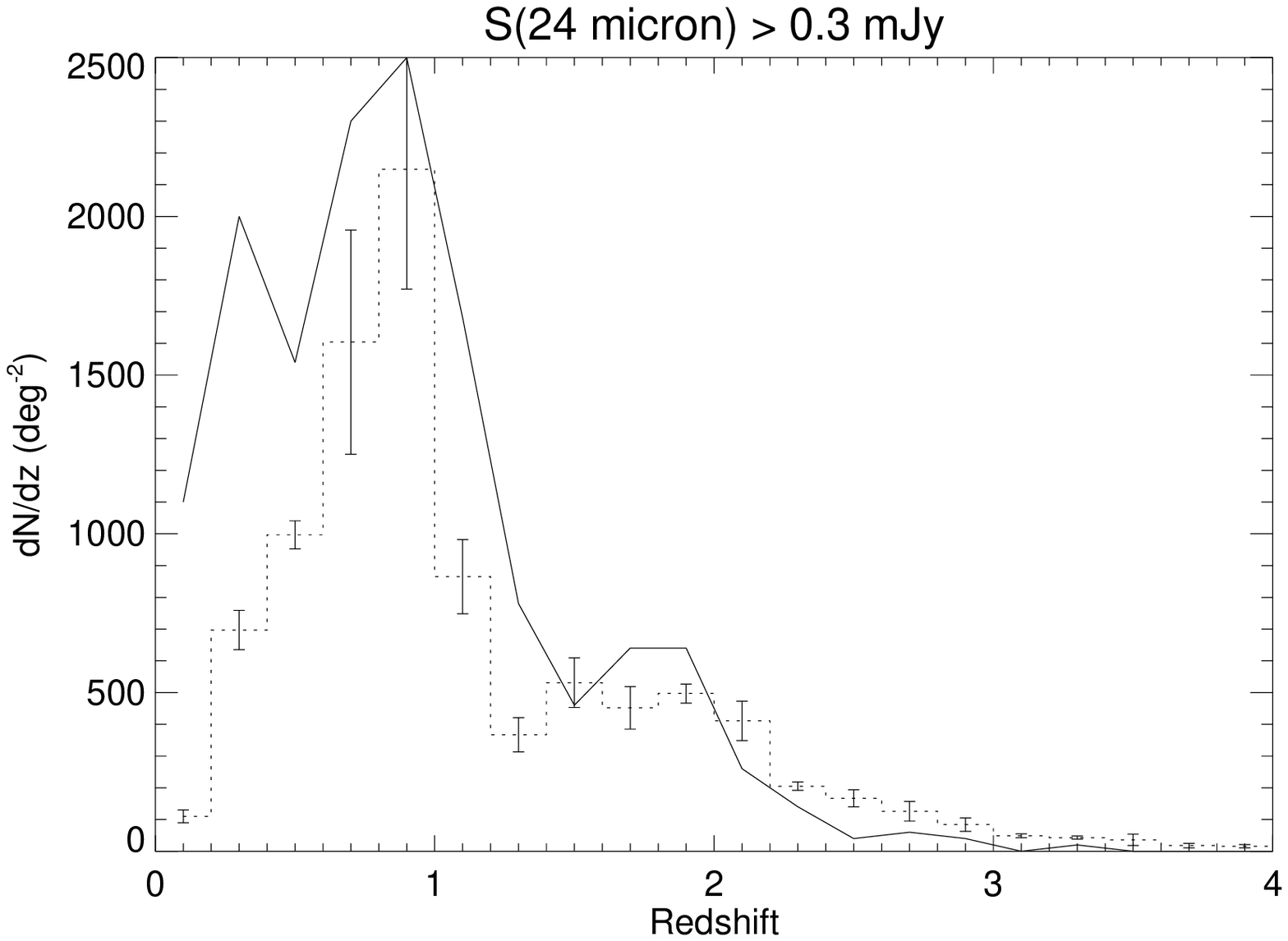}
\caption{\normalsize  Comparison of the predicted 24\micron~redshift distributions for the final model of section 4.2 (solid lines) in the central 0.25~deg$^{2}$ of the simulation with those measured in a 1.68 deg$^{2}$ region of the COSMOS field by Le Floc'h et al.~(2009). The latter authors excluded galaxies brighter than $I_{\rm{AB}}=20$~mag to minimize cosmic variance 
fluctuations at $z<0.4$. }
\label{fig:24micZDIST}
\end{figure*}

Having achieved reasonable agreement with the {\em Spitzer} MIPS source counts, particularly at the faint end of the observed flux range, we now compare the redshift distributions. In Fig.~\ref{fig:24micZDIST}, we show them for 24\micron~sources above flux limits of 0.08, 0.15 and 0.3~mJy. The observational data were taken from Le Floc'h et al.~(2009) and are based on photometric redshift analysis of 30,000 sources in a $\sim 1.68$~deg$^{2}$ region of the COSMOS field. The simulated distributions were derived from a 0.25~deg$^{2}$ region and are in good agreement with the observations. We note, however, that Le Floc'h et al.~excluded galaxies brighter than $I_{\rm{AB}}=20$~mag to minimise cosmic variance fluctuations at low redshift. As they also remarked, the small cosmic volume covered by the observations below $z \sim 0.4$ prevents any useful comparison in this regime.

\begin{figure}
\includegraphics[width=0.47\textwidth,angle=0]{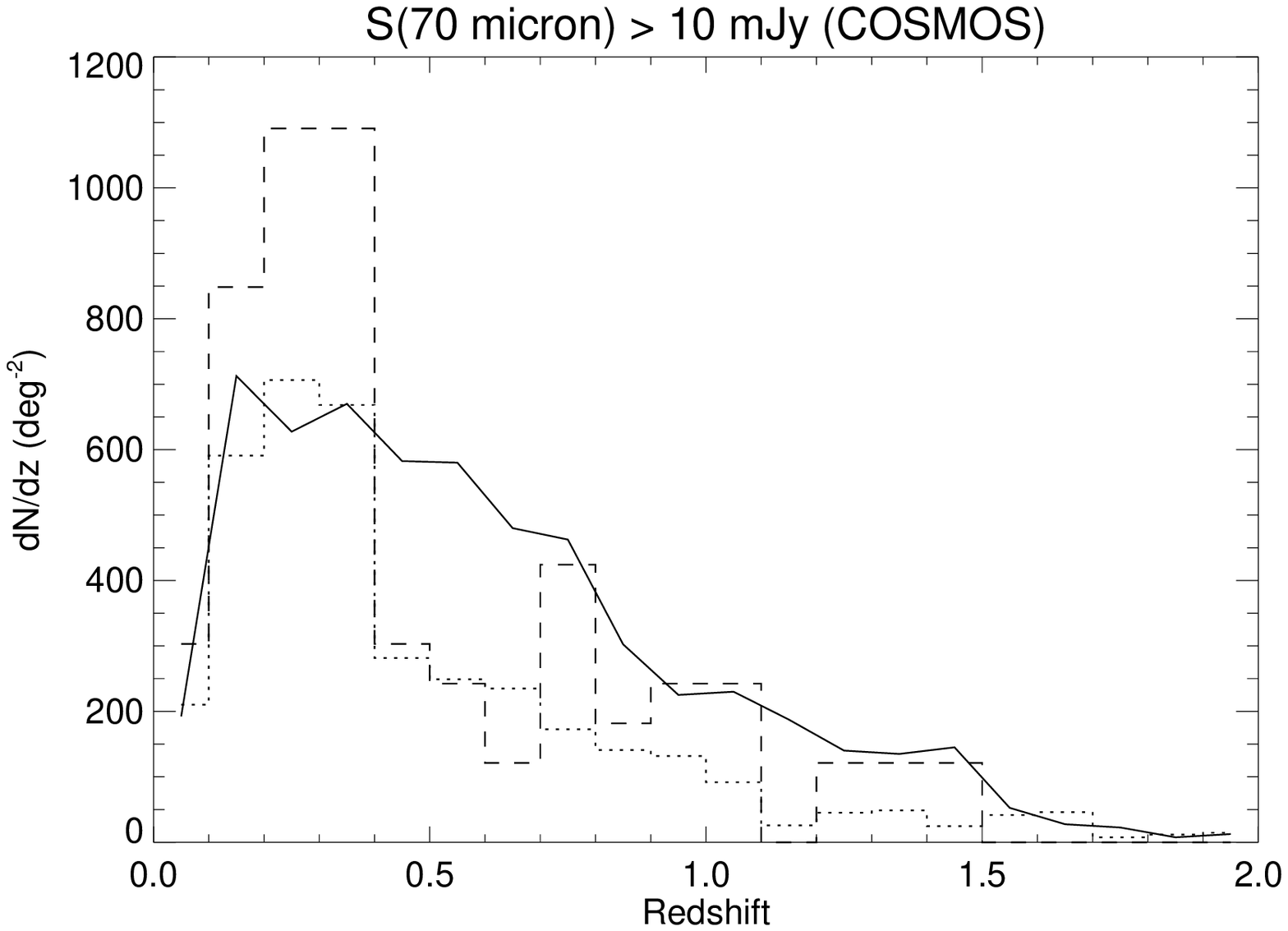}
\includegraphics[width=0.47\textwidth,angle=0]{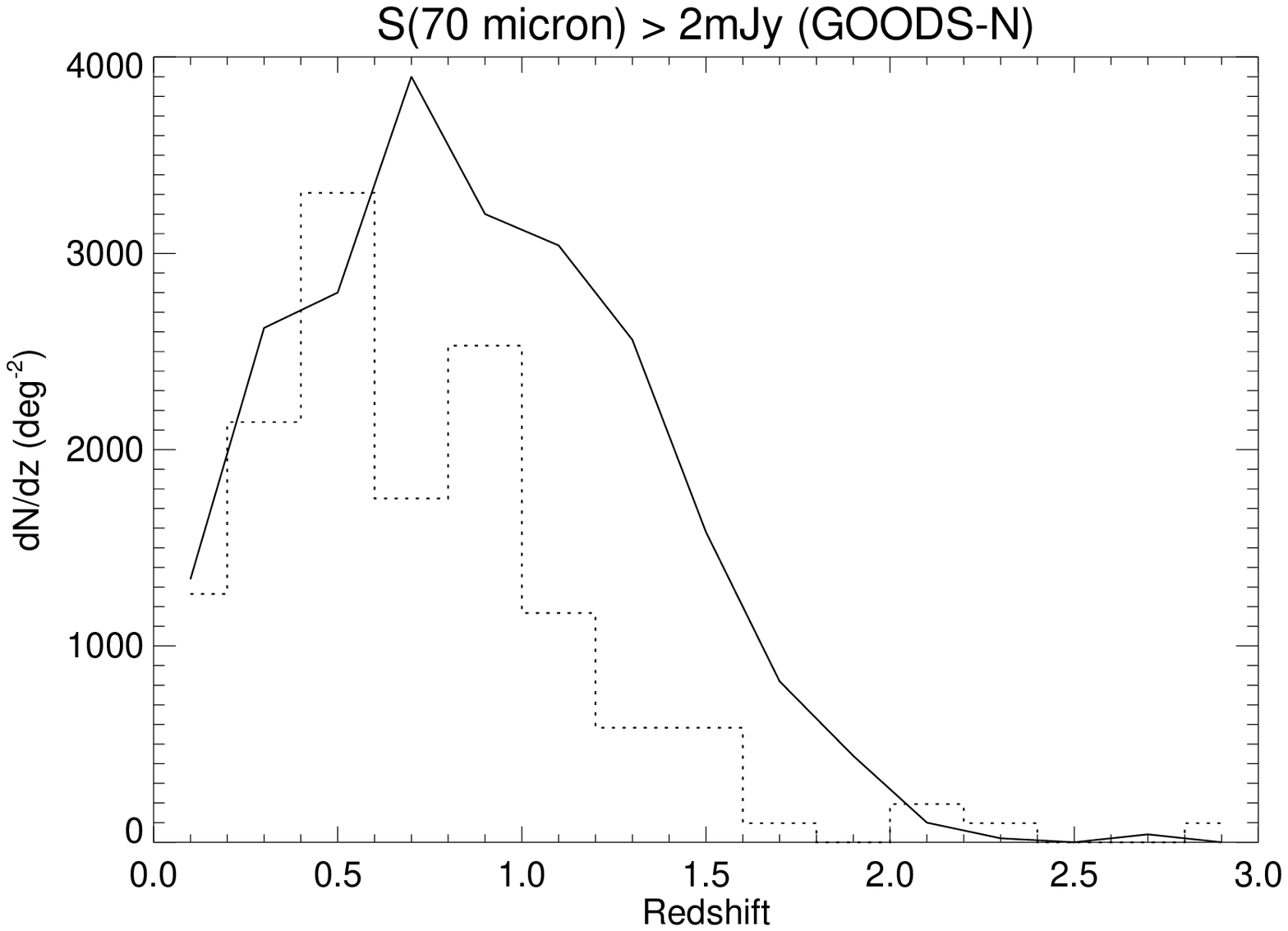}
\caption{\normalsize  Comparison of the simulated redshift distribution for the final model of section 4.2 (solid lines) with observations at 70\micron.
Upper panel: results from the COSMOS field from Hickey et al.~(in prep) for S(70\micron)$>10$\mJy~(dashed and dotted histograms are for the deep and shallow fields, respectively -- see 
section 4.3 for details); lower panel: the observed data (histogram) are the raw redshift distribution of Huynh et al.~(2007, Fig.~1) for the GOODS-N field above 2\mJy, for which the 70\micron~source catalogue is, 
however, only about 50 per cent complete.}
\label{fig:70micZDIST}
\end{figure}

At 70\micron~we compare with results from the COSMOS and GOODS-N fields (Fig.~\ref{fig:70micZDIST}) using a 4~deg$^{2}$ simulation area. The COSMOS results are based on photometric redshift analysis by Hickey et al.~(in prep.) for a shallow field covering 1.43~deg$^{2}$ and a deep sub-region of 0.165~deg$^{2}$. The deep field is complete to 10\mJy, but the shallow field is
only 60~per cent complete at this flux level; in calculating dn/dz for the shallow field we compensated for this by weighting each source by $1/c(f)$, where $c(f)$ is the completeness level at its flux, as read off from Fig.~3 of Frayer et al.~(2009). The shallow field dn/dz agrees better with the simulation at $z<0.4$, whilst at $z>0.75$ the match to the deeper field is more favourable. Redshift distributions for 70\micron~COSMOS sources have also been compiled by Kartaltepe et al.~(2009).

At a deeper flux level of 2\mJy, we compare with the raw redshift distribution derived by Huynh et al.~(2007) for the 0.05~deg$^{2}$ GOODS-N field, using a 0.25~deg$^{2}$ simulation area; there is again a deficit of observed sources but the GOODS-N source catalogue is only $\sim 50$~per cent complete at the flux limit (see Frayer et al.~2006a, Table 1), and the raw redshift distribution of Huynh et al.~(2007; their Fig.~1) appears not to correct for this incompleteness. The effects of cosmic variance for the small field observed may also be important.

At 850\micron, the redshift distributions for flux limits of a few \mJy~(Fig.~\ref{fig:TEMPEVdsc850}) have median redshifts at $z = 2-3$ and high-redshift tails extending beyond $z=4$. Thus, even though the final model has eliminated the starburst (high-L) population at $z>1.5$, it is nevertheless able to reproduce the population of sub-mm (SCUBA) galaxies which reside on average at $z \simeq 2.3$, as deduced from spectroscopic (Chapman et al.~2005) and radio-mm-FIR photometric measurements (Aretxaga et al.~2007). There is also observational evidence that the distribution extends to redshift $z=4$ and beyond (Coppin et al.~2009). This reinforces our earlier statement that the terms `normal galaxy' and `starburst', which we use to label the two Schechter function components of the $z=0$ star-forming galaxy luminosity function, should not be invested with much physical significance beyond the local Universe.

\subsection{Possible evolution in the far-infrared--radio correlation?}
\begin{figure}
\includegraphics[width=0.47\textwidth,angle=0]{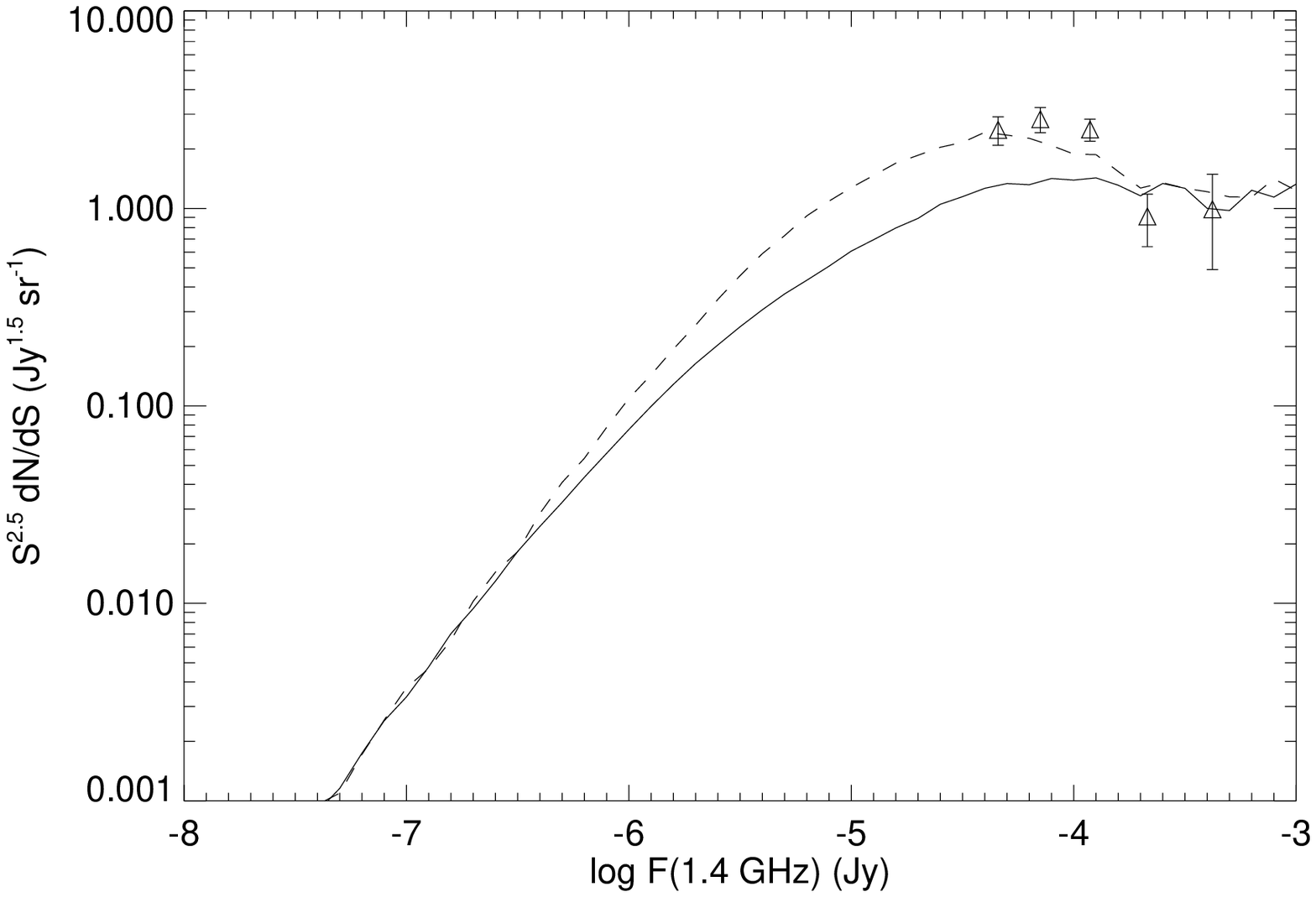}
\caption{\normalsize Differential 1.4~GHz source counts for the star-forming galaxies in the simulation with (dashed line) and without (continuous line) the 
application of the luminosity boost introduced in section 4.1. The observational data points are the star-forming galaxy source counts measured by Seymour et al.~(2008).}
\label{fig:RadioSCcheck}
\end{figure}

{\em Our default assumption is that the luminosity boost introduced in section 4.1 applies equally to the radio and infrared luminosities and that it thus reflects a 
modification to the cosmological evolution of the star-forming galaxy population as a whole.} Indeed, as we show in Fig.~\ref{fig:RadioSCcheck}, the 1.4~GHz source counts for the 
star-forming galaxies come into better agreement with the observations after the application of the luminosity boost. The relevant observations are taken from Seymour et al.~(2008) and provide a break-down of the 1.4~GHz sources counts into contributions from AGN and star-forming galaxies, with the latter dominating below 0.1~mJy. With reference to the radio source counts in Fig.~4 of W08, the increase in the star-forming galaxy contribution brings the total radio source count at 1.4~GHz into better agreement with the observations at this flux level. In light of this, the luminosity boost should be applied when using the simulated radio catalogues, as described in Appendix A.
 
Nevertheless, we cannot exclude the possibility that some contribution to $\Delta log L(L,z)$ may arise from redshift evolution and non-linearity in the far-infrared:radio correlation. 
Observational evidence for redshift evolution is at present limited, as discussed in section 3, but strong upward evolution in the relation is expected at $z>3$ due to increased inverse Compton scattering off the cosmic microwave background (e.g. Murphy~2009). In the local Universe, there is evidence that the relation may become non-linear at low luminosities (e.g. Condon~1992; Yun, Reddy \& Condon~1992; Best et al.~2005). Best et al.~(2005) found that below log L(1.4~GHz) = 22.5~W~Hz$^{-1}$ the measured local radio luminosity function falls below that derived in the far-infrared. Such non-linearity could partially account for the luminosity boosting, but it would need to evolve strongly with redshift in order to match that required in Fig.~\ref{fig:lumboost}. 

\begin{figure}
\includegraphics[width=0.47\textwidth,angle=0]{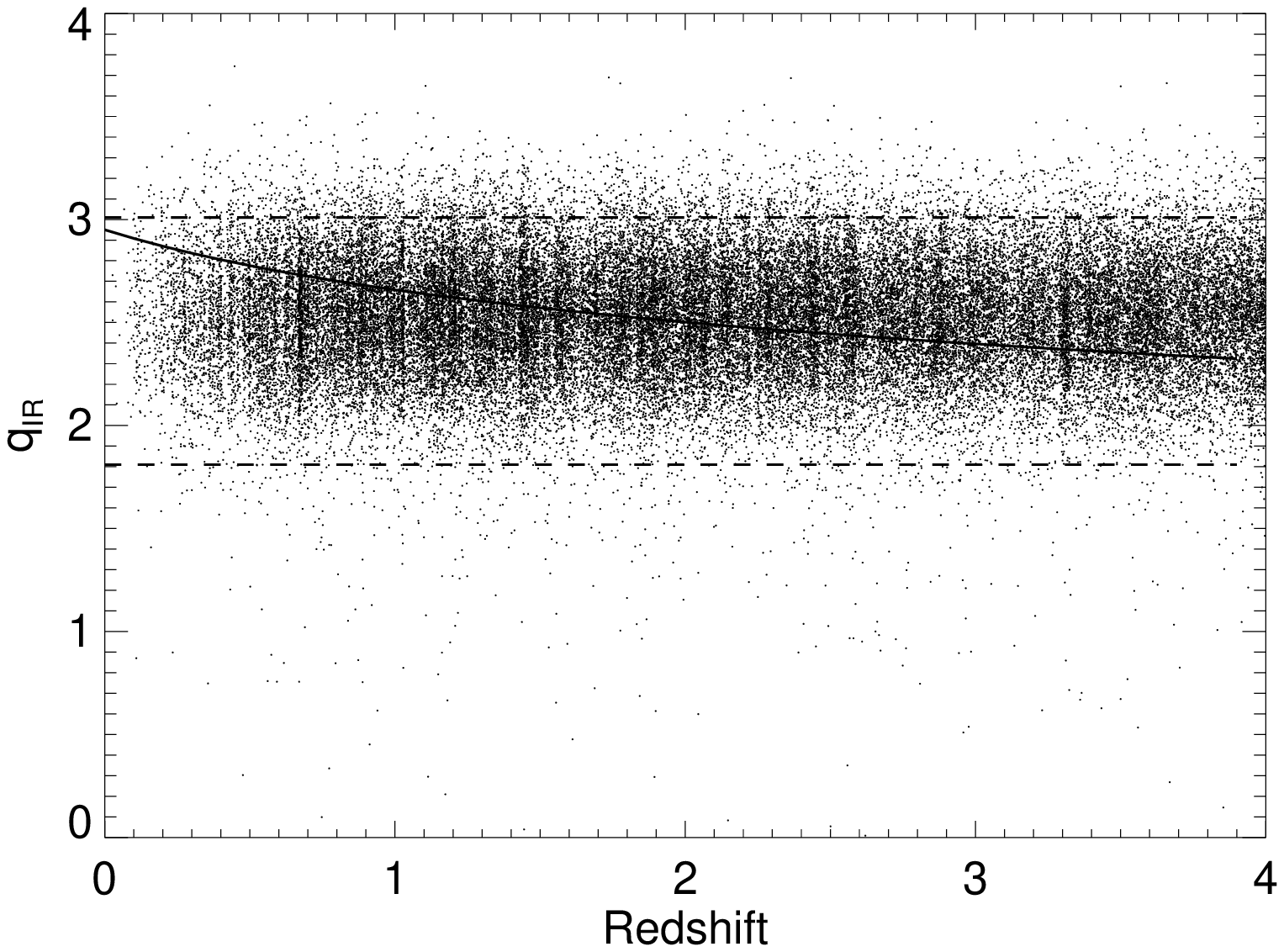}
\includegraphics[width=0.47\textwidth,angle=0]{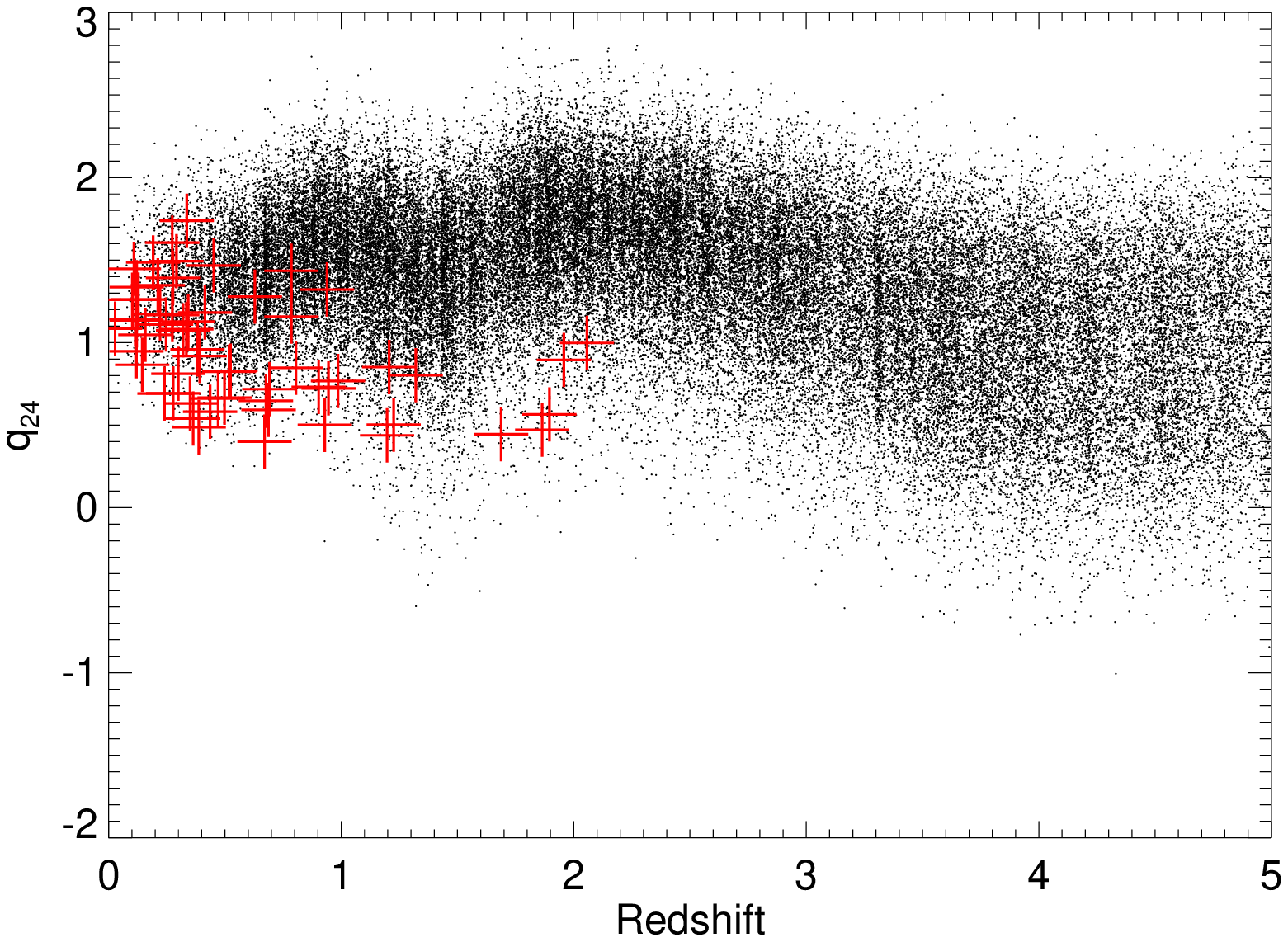}
\includegraphics[width=0.47\textwidth,angle=0]{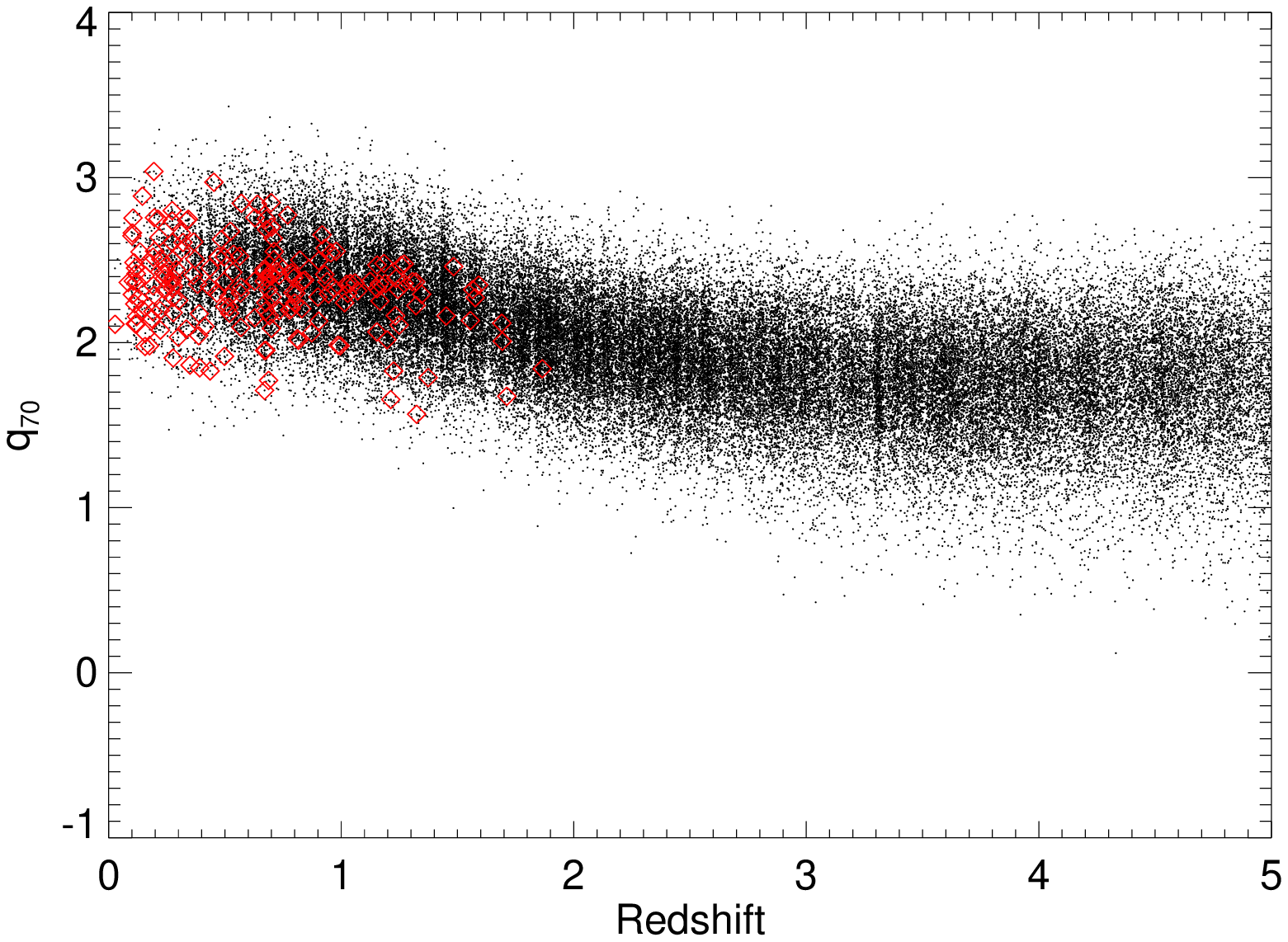}
\caption{\normalsize Upper panel: the bolometric (8--1000\micron) $q_{\rm{IR}}$ for the simulated star-forming galaxies. The dashed lines represent the $\pm 3 \sigma$ range measured by Ivison 
et al.~(2009) for a 250\micron~sample. The continuous line shows the tentative evolution measured by Ivison et al.~from stacking measurements. Middle and lower panels: $q_{\rm{24}}$ and $q_{\rm{70}}$ 
for the final model, with symbols defined as in Fig.~\ref{fig:BASELINEqparams}. }
\label{fig:qIvison}
\end{figure}

At higher redshift, major contributions in this area are expected from {\em Herschel} surveys, as suggested by initial results from {\em BLAST} (Ivison et al.~2009). To compare with the latter
we show in Fig.~\ref{fig:qIvison} the simulated quantity $q_{\rm{IR}}$, defined analogously to eqn.~(1) but with the far-infrared luminosity replaced by the rest-frame 8--1000\micron~luminosity, L(TIR), 
(eqn.~4 of Ivison et al.). The simulated values are in good accord with the value of $q_{\rm{IR}}=2.41 \pm 0.2$ for the 250\micron-selected sample of Ivison et al.~(2009), with no evidence of 
redshift dependence. Ivison et al.~also determined $q_{\rm{IR}}$ for a sample derived by stacking at the positions of 24\micron~sources and found tentative evidence for a decline of the 
form $(1+z)^{-0.15 \pm 0.03}$, but we do not attempt to replicate the selection effects associated with the definition of this particular sample. The lack of redshift dependence in the 
simulated $q_{\rm{IR}}$ primarily reflects the redshift-invariance we assumed for the far-infrared--radio relation in eqn.~(1), in conjunction with the only mildly non-linear dependence of L(TIR) on 
L(FIR) in Fig.~\ref{fig:LTIRtoFIR_Rieke}.

For completeness, we also show in Fig.~\ref{fig:qIvison} the quantities $q_{\rm{24}}$ and $q_{\rm{70}}$ as functions of redshift, for comparison with those of the baseline model shown in 
Fig.~\ref{fig:BASELINEqparams}. Reflecting the imposed template evolution, $q_{\rm{24}}$ increases more strongly with redshift out to $z=1$ than in the baseline model, whilst $q_{\rm{70}}$ remains 
effectively constant out to $z=1$.

\section{PREDICTIONS FOR HERSCHEL SURVEYS}

\begin{figure}
\includegraphics[width=0.47\textwidth,angle=0]{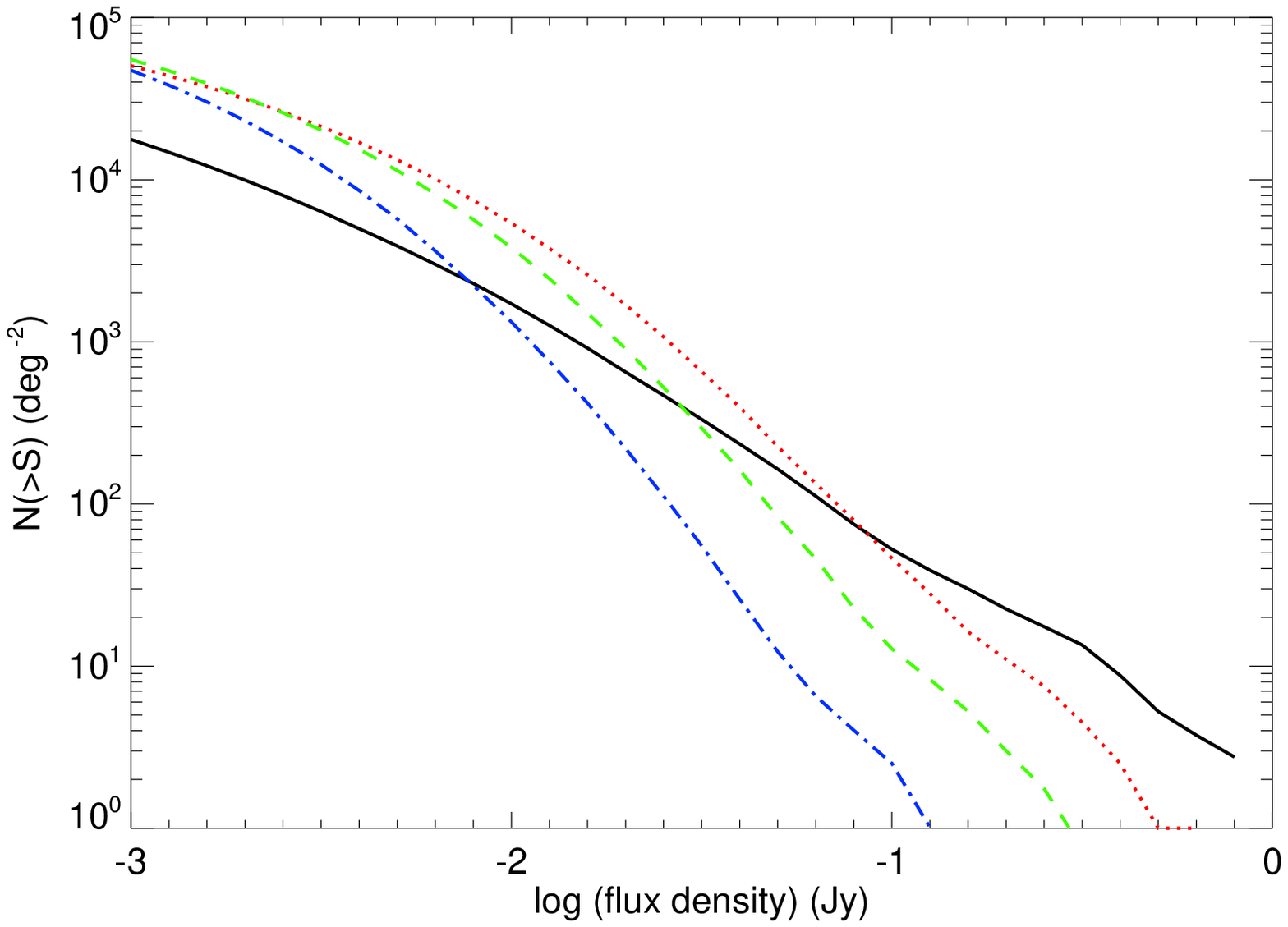}
\includegraphics[width=0.47\textwidth,angle=0]{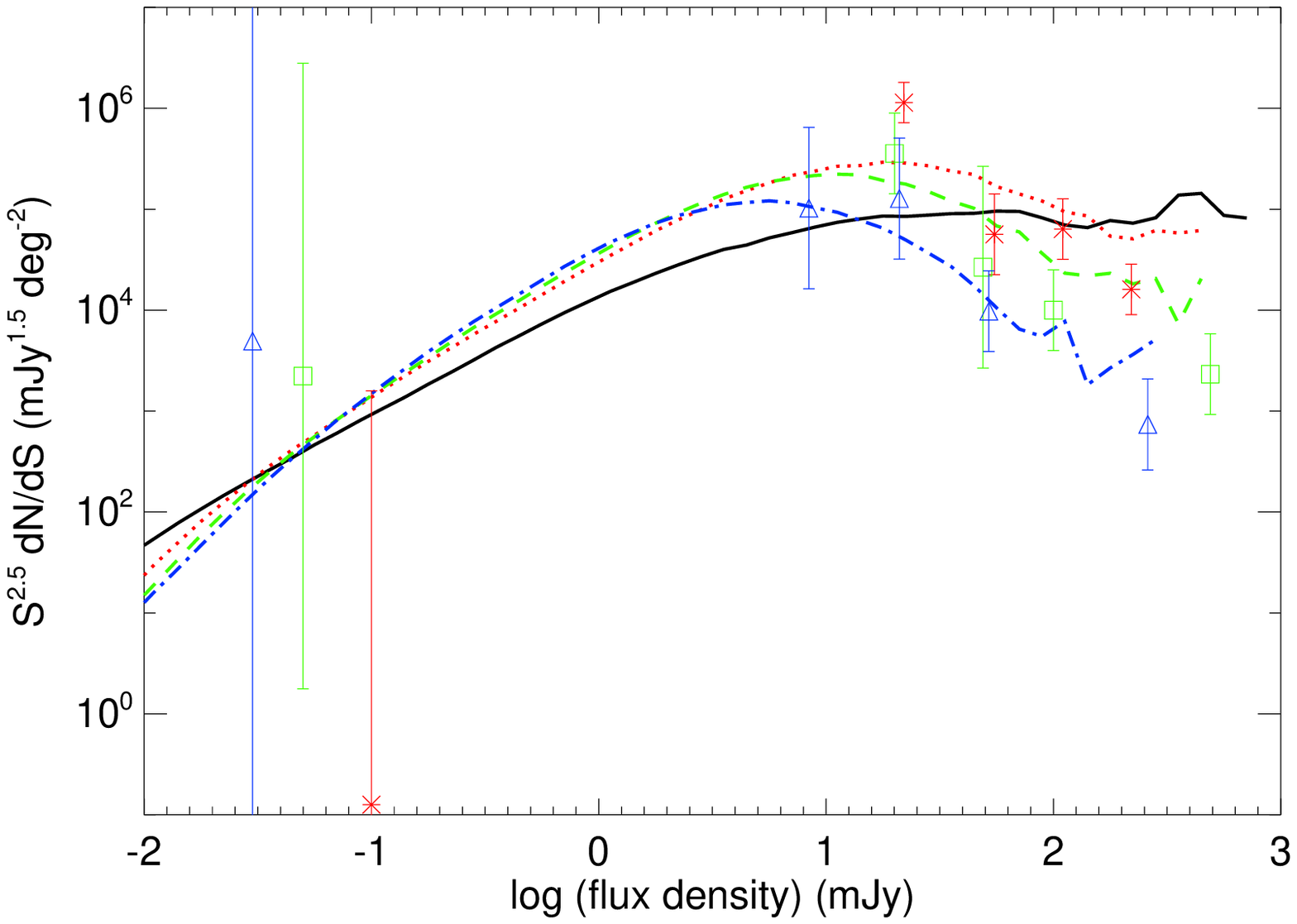}
\caption{\normalsize  Upper panel: predicted {\em Herschel} integral source counts at 100~(black solid line), 250~(red dotted line), 350~(green dashed) and 500\micron~(blue dot-dashed line); lower panel: in differential format for comparison with the {\em BLAST} counts from Table S2 of Devlin et al.~(2009), which are shown by the points with error bars (asterisks, 250\micron; squares, 350\micron; triangles, 500\micron).}
\label{fig:Herschel_SC}
\end{figure}

\begin{figure}
\includegraphics[width=0.47\textwidth,angle=0]{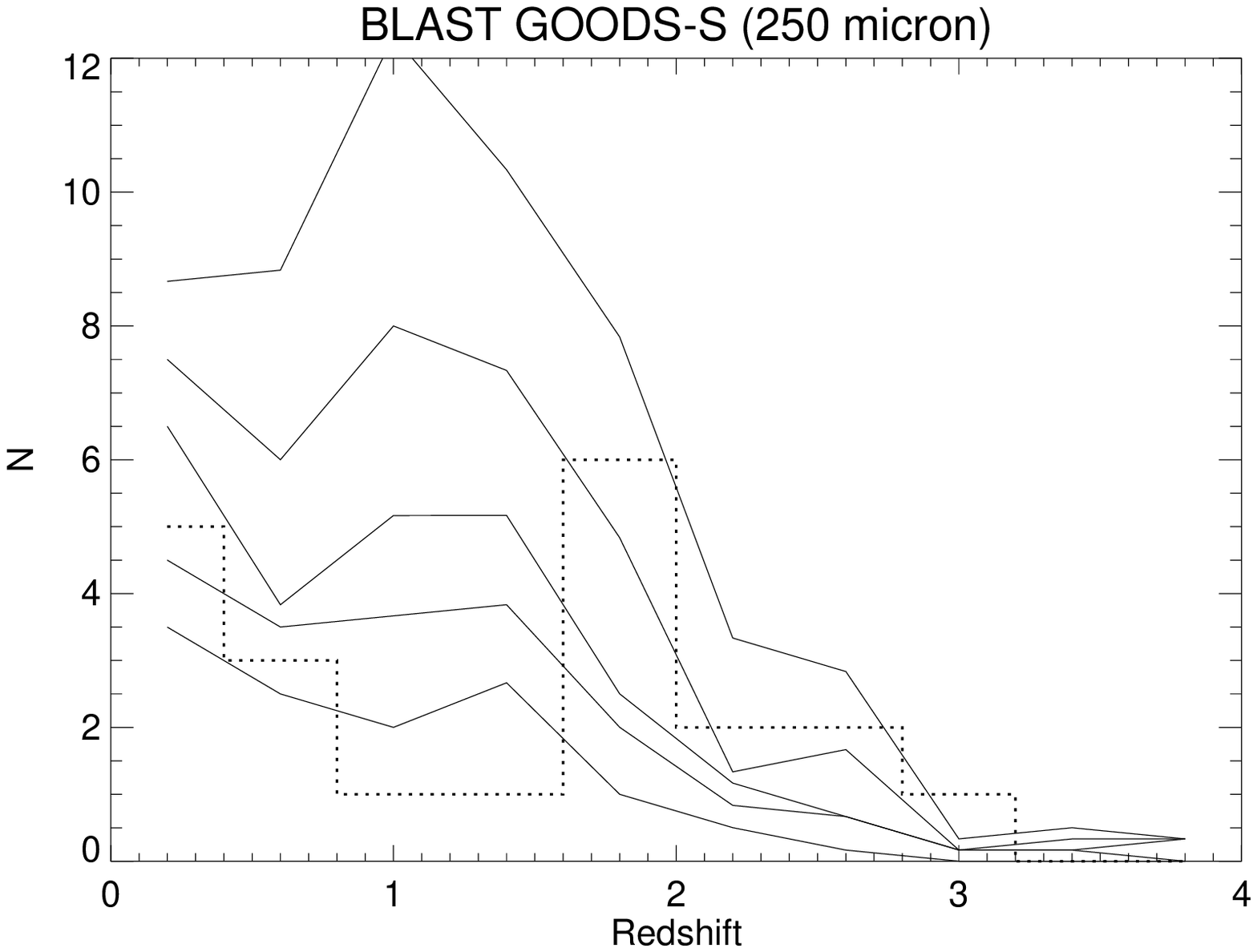}
\caption{\normalsize A comparison of the redshift distribution of {\em BLAST} 250\micron~sources brighter than 35\mJy~in GOODS-S (Dunlop et al.~2009) (histogram) with the simulated distributions for flux limits of 
25, 30, 35, 40 and 50\mJy~(solid lines, top to bottom), determined from a 0.25~deg$^{2}$ simulation area.}
\label{fig:BLAST250zdist}
\end{figure}

We now present some predictions for forthcoming {\em Herschel} surveys and a comparison with results from the {\em BLAST} mission which flew
a prototype of the SPIRE instrument on a telescope half {\em Herschel}'s size. In Fig.~\ref{fig:Herschel_SC}, we show predicted integral source counts for surveys at 100, 250, 350
and 500\micron~which correspond closely with those predicted by the backward-evolution model of Valiante et al.~(2009). Also shown in this figure are normalised differential source 
counts at 250, 350 and 500\micron, which compare reasonably well with the {\em BLAST} counts as determined from the `P(D)' analysis of Devlin et al.~(2009) (see also Patanchon et al.~2009). At 250 and 350\micron, the model marginally exceeds the {\em BLAST} measurements in the highest flux bins, as also hinted at in the 160\micron~comparison above $\sim 100$\mJy. In Fig.~\ref{fig:BLAST250zdist} we compare with the `complete' redshift distribution derived by Dunlop et al.~(2009) for the 150~arcmin$^{2}$~GOODS-S field at 250\micron; the observed
sample consists of 20 sources with redshifts down to a nominal flux limit of 35\mJy, although small number statistics and cosmic variance limit the scope of the comparison.

Although the $S^{3}$ online interface allows the user to query the database to generate redshift distributions for arbitrary radio and 
infrared flux cuts and survey areas, we follow Lacey et al.~(2009) and show in Fig.~16 dn/dz plots for a selection of 
{\em Herschel} Key Programme\footnote{http://herschel.esac.esa.int/Key\underline{  }Programmes.shtml} surveys: (i) the ultra-deep 0.012~deg$^{2}$~`pencil beam' GOODS-{\em Herschel} field\footnote{http://herschel.esac.esa.int/Docs/KPOT}; (ii) Levels 1 and 5 of the {\em Herschel} Multi-tiered Extragalactic Survey (HERMES \footnote{http://astronomy.sussex.ac.uk/$\sim$sjo/Hermes}); (iii) the wide ($\sim 600$~deg$^{2}$) but shallow ATLAS survey (Eales et al.~2009). Table~1 shows their area coverages, multi-band sensitivities, predicted numbers of galaxies and the fractions at $z>1$ and $z>2$. Our dn/dz predictions were derived from a $0.25$~deg$^{2}$ simulation area and scaled to the required survey areas, except for the ATLAS and HERMES Level 5 predictions for which we used a 4~deg$^{2}$ area . The predicted redshift distributions correspond quite well with those of Lacey et al.~(2009) (L09), despite the differences in our simulation methodologies. For GOODS-{\em Herschel}, our predicted 
redshift distributions exhibit broader peaks than those of L09, extending beyond $z=1$; similarly, for HERMES Level 1 and 5 at 250\micron~our distribution peaks just above $z=1$ and that of L09 just below. At 100 and 160\micron~in HERMES Level 5, we predict broad redshift distribution with a tail extending beyond $z=1$ whereas L09 predict distributions strongly peaked at $z=0.2$. For the ATLAS survey, the distributions of L09 exhibit marginally higher peaks than ours at $z \leq 0.2$; the L09 distributions cut-off sharply above $z \sim 1.5$, as do ours, due to the step function cut-off in the starburst population above $z=1.5$.

\begin{table*}
\caption{A selection of {\em Herschel} Key Programme surveys for which we provide dn/dz plots in section 5.}
\begin{tabular}{|lllllll|}\hline
Survey                & Area           & Band          & Depth & Total$^{\star}$ & f(z$>$1)$\dagger$ & f(z$>$2)$\dagger\dagger$ \\
                      & (deg$^{2}$)    & (\micron)     & (\mJy) &  &  & \\ \hline
GOODS-{\em Herschel}  & 0.012          & 100           & 0.6    & 320  & 0.61 & 0.14 \\
Ultra-deep            &                & 160           & 0.9    & 460 & 0.69 & 0.25 \\
Hermes Level 1        & 0.11           & 250           & 4.2    & 1850 & 0.69 & 0.24 \\
                      &                & 350           & 5.7    & 1080 & 0.75 & 0.29 \\
                      &                & 500           & 4.9    & 680 & 0.82 & 0.37 \\
Hermes Level 5        & 27             & 100           & 27     & 11440 & 0.24 & 0.0018  \\
                      &                 & 160           & 39  & 13420   & 0.41 & 0.033 \\
                      &                 & 250           & 14   & 85700 & 0.63 & 0.18 \\
                      &                & 350           & 19    & 27000 & 0.69 & 0.24 \\
                      &                & 500           & 16    & 10470 & 0.79 & 0.34 \\
ATLAS                 & 600             & 100           & 67   & 59700 & 0.15 & 1.6E-04   \\
                      &                 & 160           & 94   & 57170 & 0.30 & 8.3E-03   \\
                      &                 & 250           & 46   & 165100 & 0.51 & 0.10 \\
                      &                & 350           & 62    & 28530 & 0.51 & 0.12 \\
                      &                & 500           & 53    & 5720 & 0.63 & 0.19 \\ \hline
\end{tabular} \\
$\star$ Total number of simulated galaxies in survey area at $z=1-4$ \\
$\dagger$ Fraction at $z>1$ \\
$\dagger\dagger$ Fraction at $z>2$ \\
\end{table*}

\begin{figure}
\begin{centering}
\includegraphics[width=7.8cm,angle=0]{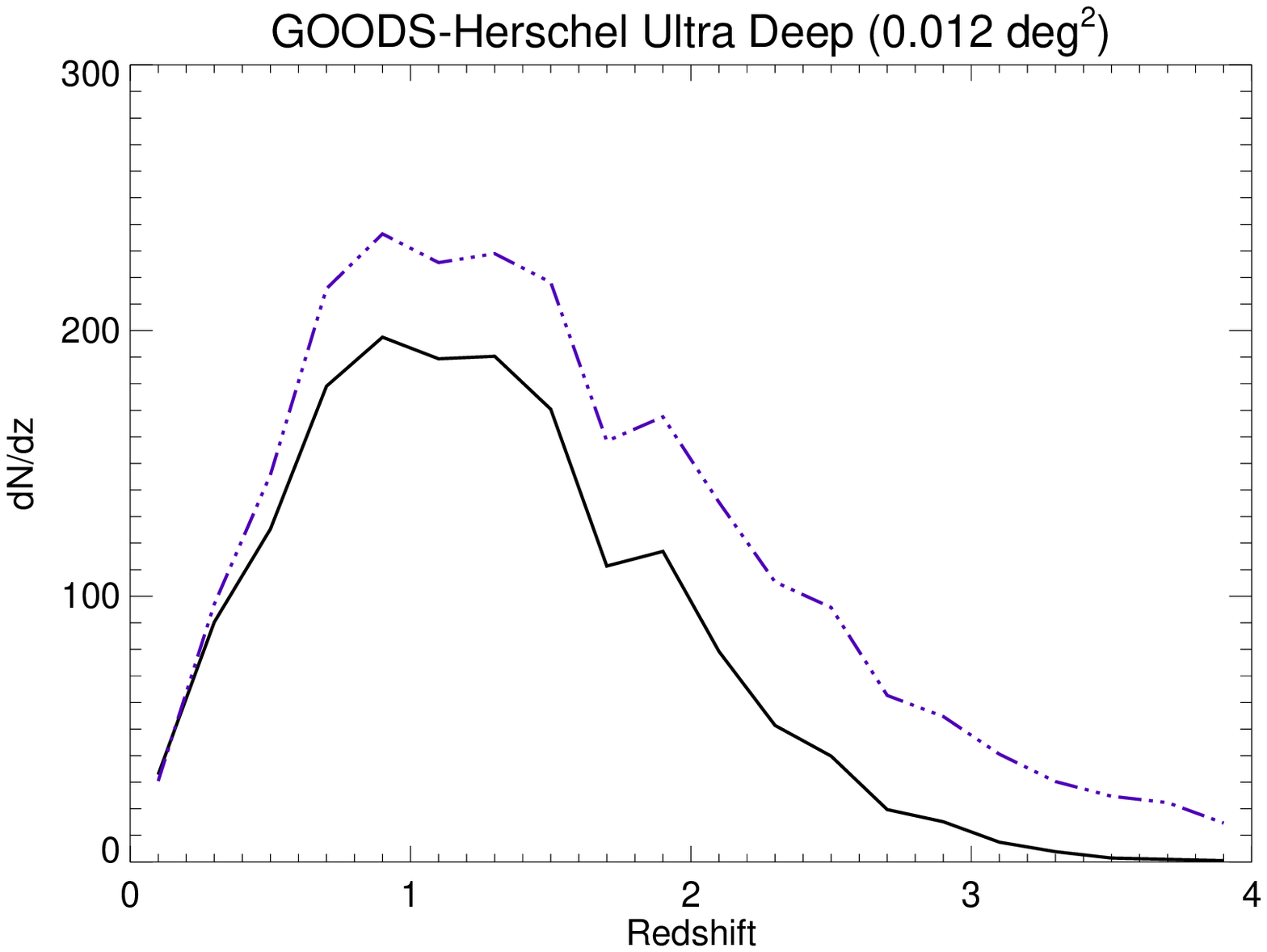}
\includegraphics[width=7.8cm,angle=0]{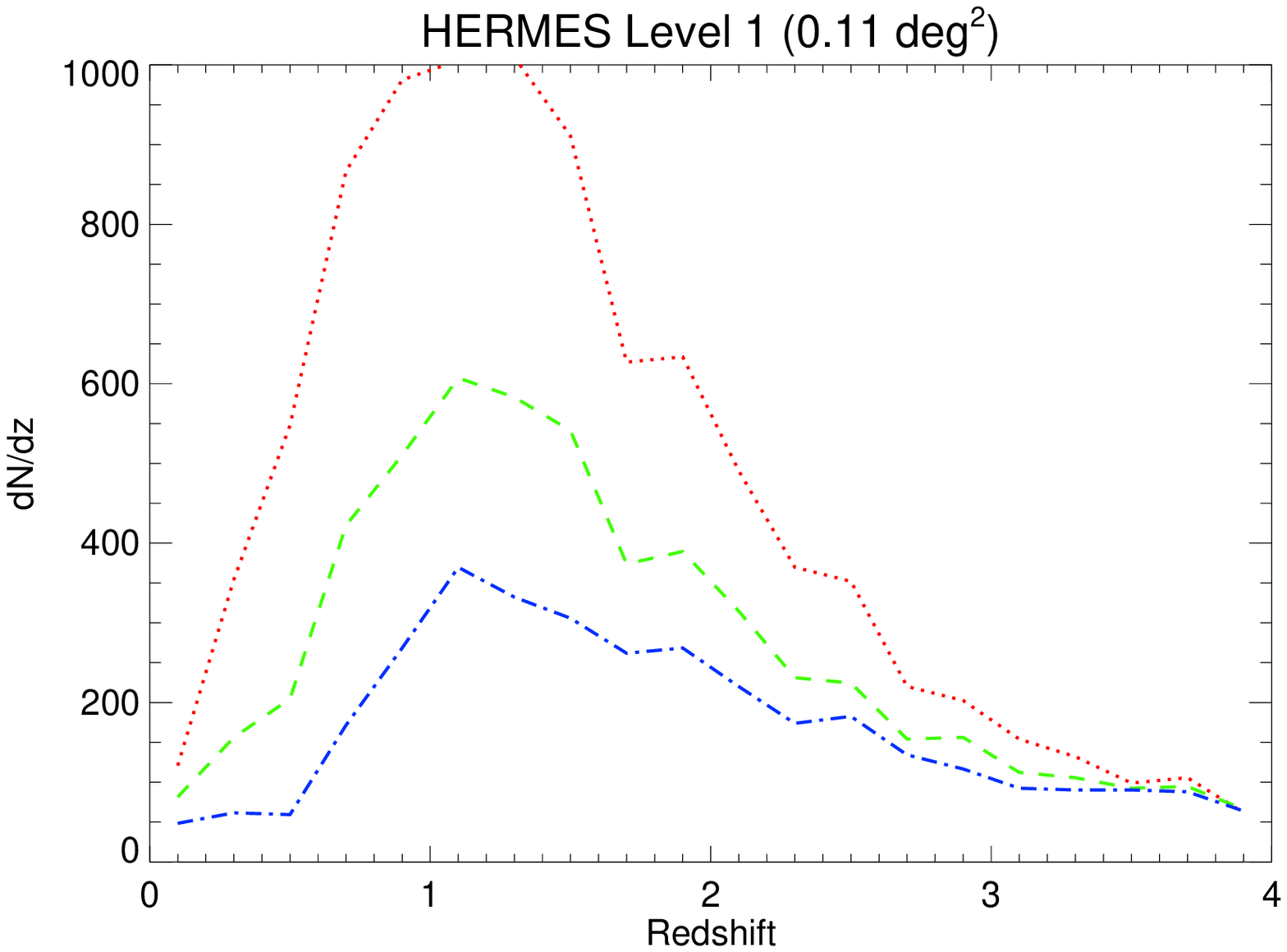}
\includegraphics[width=7.8cm,angle=0]{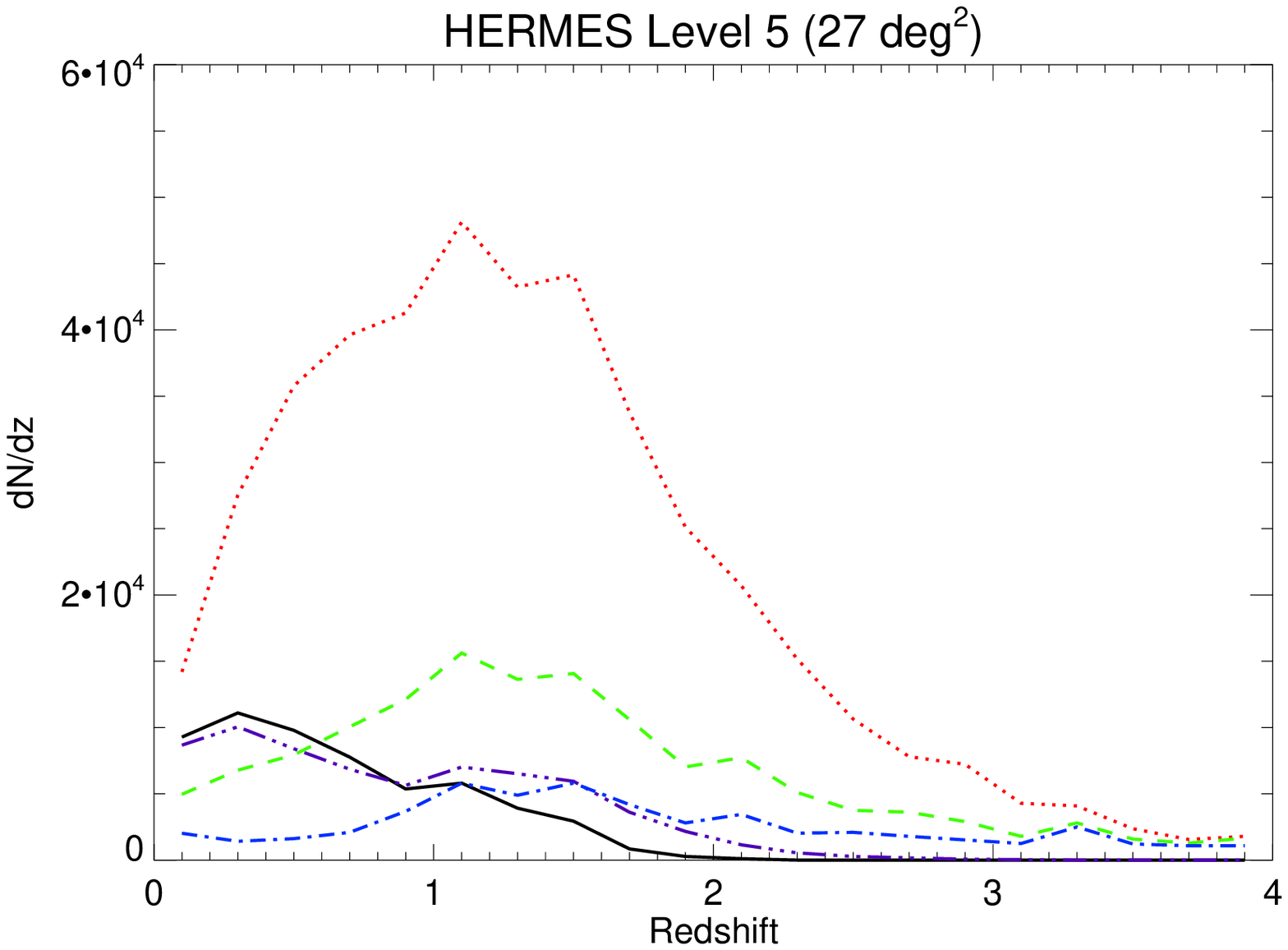}
\includegraphics[width=7.8cm,angle=0]{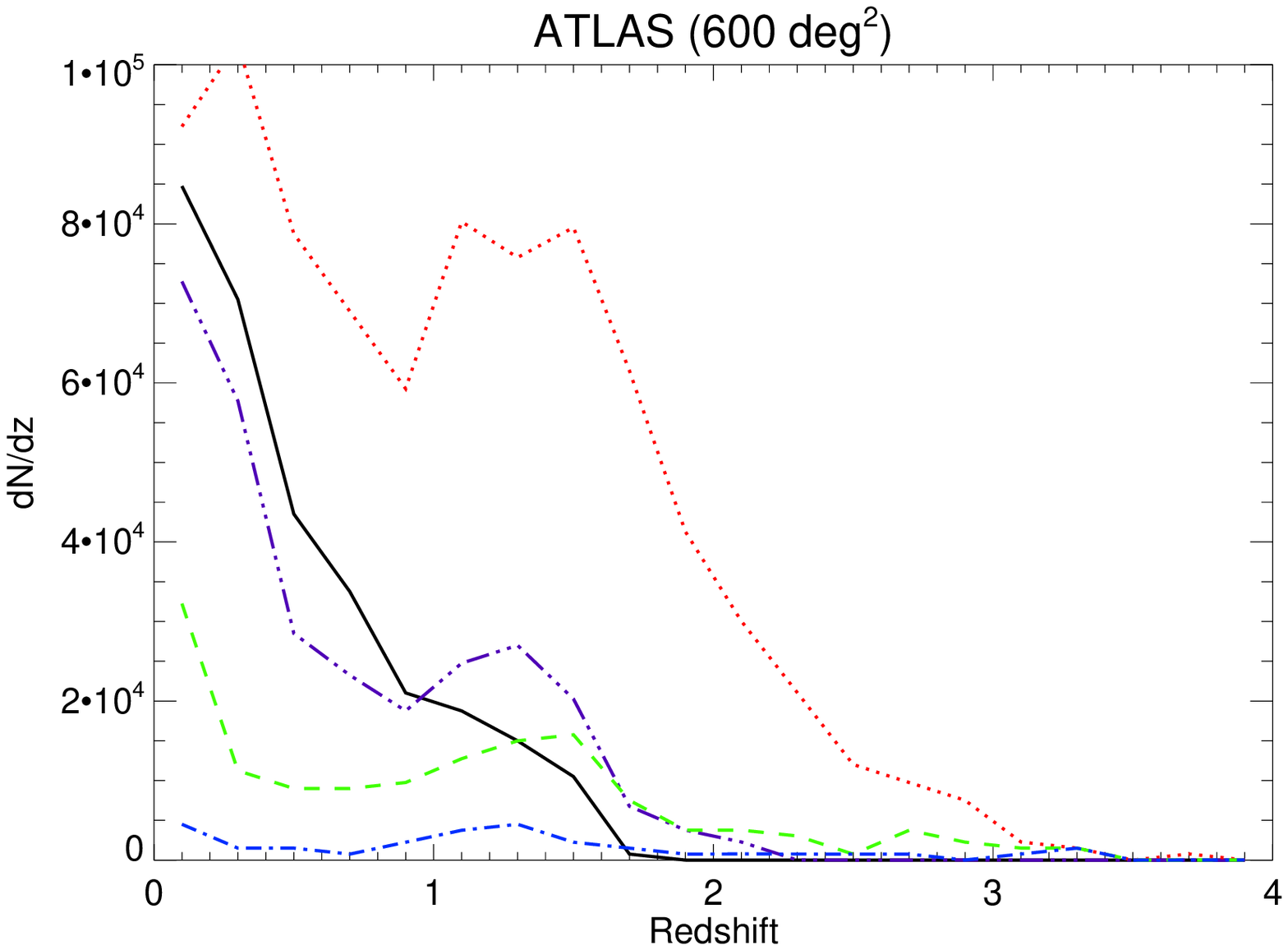}
\caption{\normalsize Predicted redshift distributions for a selection of {\em Herschel} Key Programme surveys (listed in Table 1) at 100\micron~(black solid lines), 160\micron~(purple dot-dot-dashed lines), 250\micron~(red dotted lines), 350\micron~(green dashed lines) and 500\micron~(blue dot-dashed lines)}
\end{centering}
\label{fig:Herschelzdist}
\end{figure}

\section{CONCLUSIONS}

We have post-processed the S$^{3}$-SEX semi-empirical simulation of the extragalactic radio continuum sky (W08) to make
predictions for {\em Herschel} surveys at far-infrared wavelengths. Existing observations in the mid-infrared with {\em Spitzer} at 24, 70 and 160\micron~and at 850\micron~with {\em SCUBA}, together impose strong constraints on the assignment of infrared SEDs to
the star-forming galaxies as a function of redshift. Our principal findings, incorporated into the final model, are as follows: 

(i) In order to match the 70\micron~counts, the star-forming galaxies are required to undergo stronger luminosity evolution than assumed by W08 for the radio simulation. 
It may be that W08 simply used an inaccurate evolutionary prescription based on the information available at that time; alternatively, it could be 
that there is genuine differential evolution between the far-infrared and radio populations, as a result of an evolving non-linearity in the far-infrared:radio correlation which is already apparent locally at low 
luminosity . Results from {\em Herschel} and {\em SKA} precursors will resolve this issue. 

(ii) From the local Universe to redshift $z=1$, star-forming galaxies are required to develop progressively cooler SED templates at fixed L(FIR) once the latter
has been set by the far-infrared--radio correlation; the rest-frame 60--100\micron~colour is assumed to evolve as $\rm{log} L_{\rm{60}}/L_{\rm{100}} \sim (1+z)^{-2.5}$; 
beyond $z=1$, this evolution must go into reverse in order not to overproduce the 850\micron~source counts. 

Our chosen model compares favourably with recent far-infrared survey results from {\em BLAST}. This inspires confidence in using it to make predictions for {\em Herschel} Key Programme surveys. Our predicted source counts and 
redshift distributions correspond closely with those of Valiante et al.~(2009) and Lacey et al.~(2009), respectively, despite the clear differences in the methodologies of our models. This suggests that the 
combination of {\em Spitzer} mid-infrared and 850\micron~data impose strong constraints on the allowable model parameter space. Data products from our simulation are available on the 
S$^{3}$ website\footnote{http://s-cubed.physics.ox.ac.uk} and we expect them to serve as a valuable resource for the interpretation of {\em Herschel} surveys.  

\section*{ACKNOWLEDGMENTS}
RJW is supported by the Square Kilometre Array Design Study. MJJ acknowledges a Research Councils UK Fellowship.

{}

\appendix

\section{Description of database format}

The simulated infrared fluxes are available from our online interactive database at http://s-cubed.physics.ox.ac.uk/s3\underline{ }sex. They are
provided as supplementary columns to the {\em Galaxies Table}, the first 16 columns of which contain the existing output from the W08 radio simulation. Each 
row of this table refers to an individual galaxy; multi-component radio sources (i.e. FRI and FRII sources) are denoted by their core position and integrated
radio fluxes. A description of the supplementary columns is given in Table A1.

In addition to the mid- and far-infrared flux densities (24--1200\micron), we also provide 2.2\micron~K-band magnitudes. For the radio-loud AGN, the K-band
magnitude is used to normalize the SED; for the radio-quiet AGN and star-forming galaxies, the K-band magnitude should be considered as schematic as no
attempt was made to model accurately the stellar population and dust extinction at rest-frame optical wavelengths, even though the SED templates extend into this regime.

The {\em RQ-AGN classification flag} applies only to the RQ-AGN and specifies whether the source is classified as unobscured, Compton-thin or Compton-thick obscured, 
or whether it is part of the `excess' population which is filtered out (as described in section 2.2). 

The {\em Space-density filter flag} indicates which sources have
been filtered out due to the imposed cut-off in the space density at high-redshift. The {\em infrared} fluxes of all filtered-out sources are set to -999. 

$\Delta$log(L,z) is the luminosity boost which has been applied to the star-forming galaxies to generate the infrared fluxes, over and above the level set by the far-infrared:radio correlation.
Note that the tabulated {\em radio} fluxes for the star-forming galaxies are the original values from W08, even though the boost may also be applicable to them, as discussed in sections 4.1 and 4.4.

\begin{table}
\caption{Supplementary columns in the S$^{3}$-SEX Galaxies Table derived from the infrared extension to the radio simulation.}
\begin{tabular}{|ll|}\hline
Column       & Attribute description   \\  \hline
[1-16         & Existing radio simulation output] \\ \hline
17           & log$_{\rm{10}}$(24\micron~flux density) [Jy]  \\ 
18           & log$_{\rm{10}}$(70\micron~flux density) [Jy]  \\ 
19           & log$_{\rm{10}}$(100\micron~flux density) [Jy]  \\
20           & log$_{\rm{10}}$(160\micron~flux density) [Jy]  \\ 
21           & log$_{\rm{10}}$(250\micron~flux density) [Jy]  \\ 
22           & log$_{\rm{10}}$(350\micron~flux density) [Jy]  \\ 
23           & log$_{\rm{10}}$(450\micron~flux density) [Jy]  \\ 
24           & log$_{\rm{10}}$(500\micron~flux density) [Jy]  \\ 
25           & log$_{\rm{10}}$(850\micron~flux density) [Jy]  \\ 
26           & log$_{\rm{10}}$(1200\micron~flux density) [Jy]  \\ 
27           & 2.2\micron~K-band magnitude  \\ 
28           & RQ-AGN classification flag $^{\star}$: \\ 
             & 1=TYPE 1 (unobscured), \\
	     & 2=TYPE 2 (Compton-thin obscured), \\
             & 3=TYPE 2 (Compton thick obscured) ,\\
	     & -1=Source excluded \\
29           & Space-density filter flag: \\ 
             & (-1= filtered out; 0=retained) \\ 
30	     & $\Delta$log(L,z)  $\dagger$ \\ \hline
\end{tabular} \\
$\star$ Applies to radio-quiet AGN only. \\
$\dagger$ Applies to star-forming galaxies only.\\
\end{table}

\end{document}